\documentclass[peerreview, draftcls, 11pt]{IEEEtran}

\usepackage[dvips]{graphics}
\usepackage[dvips]{changebar}
\usepackage{amsmath,bm}
\usepackage{rotating}
\usepackage[strings]{underscore}
\usepackage{setspace}
\usepackage{times}
\usepackage{color}
\usepackage{soul}
\usepackage{amsmath} 
\usepackage{amssymb}
\usepackage{mathrsfs}
\usepackage[strings]{underscore}
\usepackage{graphics,color}
\usepackage{epstopdf}
\usepackage{graphicx}
\usepackage{multirow}
\usepackage{ifpdf}
\usepackage{tablefootnote}
\usepackage{multirow}
\usepackage{algorithmicx}
\usepackage{algpseudocode}
\usepackage{footnote}
\usepackage{gensymb}

\makesavenoteenv{tabular}
 \makesavenoteenv{table}

\newcommand{\mb}[1]{{\mathbf #1}}

\ifCLASSOPTIONcompsoc
\usepackage[caption=false,font=normalsize,labelfont=sf,textfont=sf]{subfig}
\else
\usepackage[caption=false,font=footnotesize]{subfig}
\fi

\begin{document}

\setcounter{page}{1}
\newcounter{mytempeqncnt}
\title{Matrix Completion-Based Channel Estimation for MmWave Communication Systems With Array-Inherent Impairments}
 \author{Rui Hu, Jun Tong, Jiangtao Xi, Qinghua Guo and Yanguang Yu 
 \thanks{The authors are with the School of Electrical, Computer and Telecommunications Engineering, University of Wollongong, Wollongong, NSW 2522, Australia. Email:  rh546@uowmail.edu.au, \{jtong, jiangtao, qguo, yanguang\}@uow.edu.au.} 
 } \date{}

	\maketitle

\begin{abstract}
Hybrid massive MIMO structures with reduced hardware complexity and power consumption have been widely studied as a potential candidate for millimeter wave (mmWave) communications. 
Channel estimators that require knowledge of the array response, such as those using compressive sensing (CS) methods, may suffer from performance degradation 
when array-inherent impairments bring unknown phase errors and gain errors to the antenna elements. 
In this paper, we design matrix completion (MC)-based channel estimation schemes which are robust against the array-inherent impairments. We first design an open-loop training scheme that can sample entries from the effective channel matrix randomly and is compatible with the phase shifter-based hybrid system. Leveraging the low-rank property of the effective channel matrix, we then design a 
channel estimator based on the generalized conditional gradient (GCG) framework and the alternating minimization (AltMin) approach. The resulting estimator is immune to array-inherent impairments and can be implemented to systems with any array shapes for its independence of the array response. 
In addition, we extend our design to sample a transformed channel matrix following the concept of inductive matrix completion (IMC), which can be solved efficiently using our proposed estimator and achieve similar performance with a lower requirement of the dynamic range of the transmission power per antenna. Numerical results demonstrate the advantages of our proposed MC-based channel estimators in terms of estimation performance, computational complexity and robustness against array-inherent impairments over the orthogonal matching pursuit (OMP)-based CS channel estimator. 
\end{abstract} 


\section{Introduction}

The millimeter wave (mmWave) communication has been an attractive candidate for the 5G cellular network as it is possible to realize a gigabit-per-second data transmission rate and the mmWave device manufacturing technologies have been greatly developed during the past years \cite{Key elements 5G}. Large-scale multiple-input multiple-output (MIMO) transmission is suggested for mmWave systems to compensate for the significant signal attenuation in mmWave bands. However, a fully digital transceiver structure incurs significant {power consumption} by a large number of radio frequency (RF) chains. Phase shifter- or switch-based hybrid systems that employ only a few RF chains have generated considerable interests recently \cite{channel estimation 1}, \cite{switches or phase shifters}.  

To achieve high data transmission rates, precoders and combiners should be carefully designed. They are typically designed based on the channel state information (CSI) \cite{channel estimation 1}, which is obtained by using training and channel estimation techniques. However, employing large-scale MIMO leads to a large channel matrix. Conventional channel estimators, such as the least square (LS) estimator, demand a large number of training resources, which can be impractical for hybrid systems.
In the meantime, adopting large-scale antenna array usually needs online calibration because of the array-inherent impairments due to mutual coupling, manufacture flaws, etc \cite{measurement and calibration challenges}, \cite{array error calibration}, \cite{system concept}. Such impairments are typically time-varying, e.g., due to temperature changes or hardware aging \cite{system concept}. The actual antenna element's position may deviate from its designed position and the gains of different antenna elements may be unequal. Therefore, the array response may be severely impacted. Though online calibration methods can help compensate for the imperfections, some of them require special hardware design \cite{Cali1}, \cite{Cali2}, yielding limited implementations. Therefore, suitable mmWave channel estimators should be able to reduce the training overhead and alleviate the burden of online calibration.

Fortunately, due to the poor scattering nature at mmWave frequencies, there are only a few dominant spatial paths in the mmWave channel \cite{Key elements 5G}, \cite{measurements}, which indicates that the channel can be reconstructed by using the information of those paths. Obtaining the paths’ information may require less training resources, and thus the training overhead could be reduced. As such, the channel estimation problem can be solved by finding the AoDs (angle of departure), AoAs (angle of arrival) and path gains of the dominant paths in the channel. Compressive sensing (CS)-based channel estimators have been proposed in \cite{channel estimation 1}, \cite{channel estimation via OMP}, \cite{codebook design} to find the paths' information. One main idea of these estimators is to search for the angle pairs in a predefined dictionary based on the training information. Therefore, their performances are highly dependent on the  quality of the dictionary which is usually designed based on the array response. Also, the CS-based estimators may suffer from a heavy computational load when a high-resolution dictionary is applied to achieve better performance.
Channel estimators that achieve high-resolution estimation of AoDs and AoAs are proposed in \cite{beam pair AoA AoD}, \cite{angle_atomic}, \cite{cov_atomic}. In particular, \cite{beam pair AoA AoD} designs structured training beam pairs to achieve high-resolution estimation. In \cite{angle_atomic} and \cite{cov_atomic}, the AoAs/AoDs finding problem is formulated as an atomic norm minimization problem and solved by using semidefinite programming (SDP). These methods still rely on the knowledge of the array response to solve the problem. There are also two-stage estimators which first use methods, e.g., matrix completion \cite{MC_CS} and PARAFAC decomposition \cite{PARAFAC}, to construct a matrix containing the AoA/AoD information, and then use CS methods to find the AoA/AoD pairs. The methods at the first stage can be independent of the array response, but the CS methods at the second stage may still rely on the knowledge of the array response. However, as mentioned above, due to the presence of the array-inherent impairments, the array response of the uncalibrated arrays may not be accurately known, which would introduce errors for the estimators relying on such knowledge. 
Therefore, such estimators can be vulnerable to array-inherent impairments. For example, for the CS-based estimators, it is challenging to construct a proper basis that the mmWave channel is aligned on without knowing the array response, and thus the basis mismatch issue will arise \cite{CS with coherent dic}, \cite{basis mismatch}, which could degrade the estimation performance. 
Apart from tackling the channel estimation problem as finding the AoA/AoD pairs, \cite{subspace estimation} estimates the subspace of the mmWave channel by adopting the Arnoldi iteration technique. This method is independent of the array response but it heavily relies on channel reciprocity since it treats the downlink channel as the transpose of the uplink channel and requires closed-loop training. The channel estimation problem is solved in \cite{channel covariance} by utilizing the channel covariance matrix. Though this method is irrelevant to basis, it requires 
knowledge of the channel covariance matrix, which is difficult to obtain in practice.

In this paper, we propose an alternative channel estimation scheme leveraging the tool of matrix completion (MC). We target narrow-band mmWave channels \cite{measurements}. We focus on single-user, phase shifter-based, fully connected hybrid systems, and consider array-inherent impairments. 
We formulate the channel estimation problem as an MC problem by exploiting the low-rankness of typical mmWave channels. We then provide a training design that is compatible with the hybrid system, which involves the design of the hybrid transceivers such that the entries of the channel matrix can be properly sampled. A generalized conditional gradient (GCG) framework \cite{GCG} is applied to implement the MC-based channel estimator and an alternating minimization (AltMin) approach is introduced to accelerate the convergence of the estimation algorithm. 
Since our proposed channel estimator is independent of the array response, it can be effective even when the array is not perfectly calibrated, e.g., when there are phase errors and gain errors in the array. 
We further generalize our scheme to an inductive matrix completion (IMC) design. The resulting channel recovery problem can be solved directly by using our proposed channel estimator.   
We evaluate the performance of our proposed estimator in terms of normalized mean square error (NMSE) and spectral efficiency (SE). The simulation results show that the MC schemes are immune to the phase and gain errors of the array and have better performance in terms of SE with lower computational complexity than the OMP-based CS estimator in \cite{channel estimation via OMP}.


The paper is organized as follows. We first introduce the mmWave channel model and the fully connected hybrid structure and then discuss the channel estimation problem in Section II. In Section III, we introduce the training process of our proposed channel estimation scheme and discuss the MC-based estimation algorithm. We also generalize the design to an IMC formulation in Section III. Simulation results are given in Section IV. Section V concludes the paper.


\section{The Mmwave Channel Estimation Problem}
In this section, we first introduce the mmWave channel model as well as the hybrid system and then discuss the mmWave channel estimation problem and a typical CS-based scheme.

\subsection{MmWave Channel Model}
In this paper, we consider the downlink mmWave transmission system and assume the following small-scale fading model for the mmWave channel \cite{measurements}:
\begin{equation}
\label{mmWave H}
\mb H=\frac{1}{\sqrt{L}}\displaystyle\sum_{k=1}^K\displaystyle\sum_{l=1}^Lg_{kl}\mb a_r(\phi^{r}_{kl},\theta^{r}_{kl})\mb a_t^H(\phi^{t}_{kl},\theta^{t}_{kl}),
\end{equation}
where $K\sim\text{max}\{{\rm{Poisson}}(\lambda),1\}$ is the number of clusters with $\lambda$ as the mean of the Poisson distribution and $L$ is the number of rays within each cluster. The complex small-scale fading gain $g_{kl}$ on the $l$-th ray of the $k$-th cluster follows a complex Gaussian distribution, i.e., $g_{kl}\sim\mathcal{CN}(0,\gamma_k)$, where $\gamma_{k}$ is the fraction power of the $k$-th cluster and can be modeled  using \cite[eq (7)]{measurements}.

In this paper, we assume the uniform linear array (ULA) and the uniform square planar array (USPA).
$\mb a_{r}(\phi^{r}_{kl},\theta^{r}_{kl})$ and $\mb a_{t}(\phi^{t}_{kl},\theta^{t}_{kl})$ represent the receiving and transmitting array response vectors, respectively, where $\phi^{r}_{kl}$, $\phi^{t}_{kl}$, $\theta^{r}_{kl}$ and $\theta^{t}_{kl}$ are the azimuth AoA, the azimuth AoD, the elevation AoA and the elevation AoD on the $l$-th ray of the $k$-th cluster, respectively. Moreover, these angles are characterized by cluster center angles and ray angle shifts. Take azimuth AoA as an example: $\phi^{r}_{kl}=\phi^{r}_{k}-\varphi^{r}_{kl}$, where $\phi^{r}_{k}$ is the center angle of the $k$-th cluster and $\varphi^{r}_{kl}$ is the angle shift of the $l$-th ray away from the center angle of the cluster. Similarly, $\theta^{r}_{kl}=\theta^{r}_{k}-\vartheta^{r}_{kl}$, $\phi^{t}_{kl}=\phi^{t}_{k}-\varphi^{t}_{kl}$ and $\theta^{t}_{kl}=\theta^{t}_{k}-\vartheta^{t}_{kl}$. This representation indicates that each cluster covers a range of angles, and the angular spread characterizes the span of each cluster. 
In \cite{measurements}, channel measurements in the urban area of New York city are presented and the angular spread is shown in terms of the root-mean-square (rms) of all the measurements. 
At the carrier frequency $f_c=28$ GHz, angular spreads of $15.5^{\degree}, 6^{\degree}, 10.2^{\degree}$ and $0^{\degree}$ are reported for the azimuth AoA, the elevation AoA, the azimuth AoD and the elevation AoD, respectively. 

For an $N_a$-element ULA placed along the $y$ axis with distance $d$ between adjacent antennas, the array response is given by \cite{spatially precoding}
\begin{equation} 
\label{aMS}
\mb a(\phi_{kl})=\frac{1}{\sqrt{N_a}}[1, \mathrm{e}^{j \frac{2\pi}{\lambda_c }d\sin(\phi_{kl} )},
\cdots,\mathrm{e}^{j(N_a-1)\frac{2\pi}{\lambda_c}d\sin(\phi_{kl} )} ]^T, 
\end{equation} 
where $\lambda_c$ is the carrier wavelength and $N_a=N_t$ or $N_r$ is the number of antennas at the transmitter (BS) or the receiver (MS). 

For a $\sqrt{N_a}\times \sqrt{N_a}$ USPA placed on the $yz$ plane with distance $d_c$ between adjacent antennas, the array response \cite{AltMin} is
\begin{equation}
\label{kronA}
\mb a(\phi_{kl}, \theta_{kl})=\mb a_y(\phi_{kl}, \theta_{kl})\otimes \mb a_z(\theta_{kl}),
\end{equation}
where $\otimes$ denotes the Kronecker product,  
\[\mb a_y(\phi_{kl},\theta_{kl})=\frac{1}{N_a^{\frac{1}{4}}}[1, \mathrm{e}^{j \frac{2\pi}{\lambda_c}d_c\sin(\phi_{kl})\sin(\theta_{kl})},\]
\[\cdots,\mathrm{e}^{j(\sqrt{N_a}-1) \frac{2\pi}{\lambda_c}d_c\sin(\phi_{kl})\sin(\theta_{kl})} ]^T\]
is the array response along the $y$ axis, and 
\[\mb a_z(\theta_{kl})=\frac{1}{N_a^{\frac{1}{4}}}[1, \mathrm{e}^{j \frac{2\pi}{\lambda_c}d_c\cos(\theta_{kl})},
\cdots,\mathrm{e}^{j(\sqrt{N_a}-1)\frac{2\pi}{\lambda_c}d_c\cos(\theta_{kl})} ]^T  \] is the array response along the $z$ axis.

The resulting channel $\mb H$ is an $N_r\times N_t$ matrix. The number of clusters $K$ is usually small, e.g., $K=1, 2$, or $3$, but the number of rays $L$ in each cluster can be large, e.g., $L=20$ \cite{measurements}, which yields a large number of $KL$ paths. This suggests that $\mb H$ may have a high rank $r_{\rm ch}$. 
Let $\sigma_1>\sigma_2>\cdots> \sigma_{r_{\rm ch}}$ be the singular values of $\mb H$. We may use 
\begin{equation}
\label{energy}
p_e\overset{\Delta}{=}\frac{\sum^{r_{\rm{sub}}}_{j=1}\sigma^2_j}{\sum^{r_{\rm ch}}_{i=1}\sigma^2_i}
\end{equation}
to measure the energy captured by a rank-$r_{\rm{sub}}$ approximation of $\mb H$. It has been shown that for capturing a majority of the total energy, e.g., with $p_e=0.9, 0.95$, the required rank $r_{\rm{sub}}$ is generally much smaller than $r_{\rm ch}$ according to the measurements and simulations in \cite{measurements}. Therefore, the mmWave channel can be considered as low-rank.

\subsection{Hybrid Transceivers}

\begin{figure}
\label{hybrid_sys}
\includegraphics[width=\columnwidth]{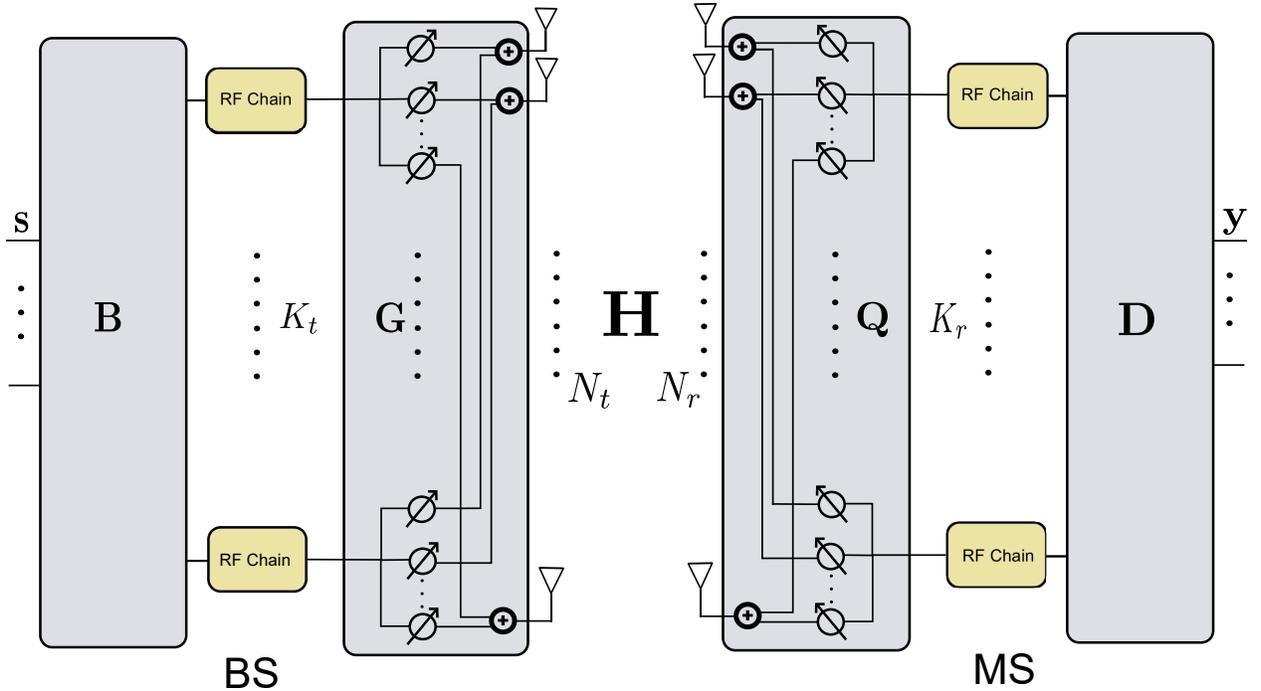}
\caption{The fully connected hybrid system}
\end{figure}

The phase shifter-based fully connected hybrid MIMO system has proven able to approximate the fully digital system in terms of SE \cite{channel estimation 1}. A point-to-point hybrid structure is shown in Fig. 1. The $N_t(N_r)$ antennas and analog phase shifters at the BS (MS) are fully connected. There are $K_tN_t$ phase shifters at the BS and $K_rN_r$ phase shifters at the MS, where $K_t\ll N_t$ and $K_r\ll N_r$ are the numbers of BS and MS RF chains, respectively. 
{For single-stream transmissions with one symbol $s$ transmitted, the received signal can be written as 
\begin{equation}
\label{received}
\mb y=\mb W^H \mb H\mb f  s +\mb W^H \mb n,
\end{equation}
where $\mb W$ and $\mb f$ are the MS receiving processing matrix and BS transmitting processing vector, respectively, and $\mb n$ is the noise vector. }
In this hybrid system, up to $K_r$ digital symbols can be received by the MS at each channel use. The traditional LS estimator, which requires at least $N_tN_r$ samples, needs at least $N_tN_r/K_r$ time slots and can be time-consuming when $K_r\ll N_r$. New methods with low sample supports may be explored to reduce the training overhead. 

\subsection{Array-Inherent Impairments}
Array-inherent impairments can cause the antenna elements' positions to deviate from their assumed ones and the gains of different antenna elements to be unequal, bringing uncertainties to the array response. 
To characterize these impairments, we use phase error $\kappa_i=2\pi\Delta_i/\lambda_c$ to represent the phase difference caused by the antenna element's position deviation $\Delta_i$, and use $\rho_i$ to denote the gain of each antenna element. With the existence of the phase error and the unequal gain effect, the array response differs from (\ref{aMS}). 

We define the gain and phase error vector at the BS or the MS as
\begin{equation}
\label{array error}
\mb e=[\rho_1\mathrm{e}^{j\kappa_1}, \rho_2\mathrm{e}^{j\kappa_2},\cdots, \rho_{N_a}\mathrm{e}^{j\kappa_{N_a}}
]^T,
\end{equation}
where $N_a=N_t$ or $N_r$. We use $\mb e_{t}$ and $\mb e_{r}$ to denote the gain and phase error vectors at the BS and MS, respectively. Let us take the MS as an example. For ULA, the actual array response is
\[ \widetilde{\mb a}_{r}(\phi^{r}_{kl})=\mb a_{r}(\phi^{r}_{kl})\odot \mb e_{r},\] where $\odot$ denotes the Hadamard product. Clearly,  
\begin{align}
\label{errorsteering}\nonumber
\widetilde{\mb a}_{r }(\phi^{r}_{kl})&=\frac{1}{\sqrt{N}}[\rho_1\mathrm{e}^{j\kappa_{1}}, \rho_2\mathrm{e}^{j (\frac{2\pi}{\lambda}d\sin(\phi^{r}_{kl})+\kappa_{2})},\\ 
&\cdots,\rho_{N_r}\mathrm{e}^{j(\frac{2\pi}{\lambda }(N_r-1)d\sin(\phi^{r}_{kl} )+\kappa_{N_r})}]^T. 
\end{align}
For USPA, 
\begin{equation}
\label{misUSPA}
\widetilde{\mb a}_{r}(\phi^{r}_{kl}, \theta^{r}_{kl}) =  \left(\mb a_{r, y}(\phi^{r}_{kl},\theta^{r}_{kl}) \otimes \mb a_{ r, z}(\theta^{r}_{kl})\right)\odot \mb e_{r}.
\end{equation}
With phase and gain errors presenting in the array, the received signal $\mb y$ in (\ref{received}) is changed to 
\begin{equation}
\label{Y_real}
\widetilde{\mb y} =\mb W^H  \mb E_{r}\mb H\mb E^H_{t}\mb f  s + \mb W^H\mb E_r\mb n,
\end{equation} 
where $\mb E_{r}$ is a diagonal matrix with $\mb e_r$ as the diagonal elements, and $\mb E_r$ is defined similarly.
The effective channel matrix $\mb H_{\rm{eff}}$ is 
\begin{equation}
\label{H_eq}
\mb H_{\rm{eff}} =\mb E_r\mb H\mb E^H_t.
\end{equation}
Note that $\mb E_r$ and $\mb E_t$ are unknown in practice.  
\subsection{A Typical CS-Based Scheme}
Channel estimation aims to recover the unknown $\mb H$ (or $\mb H_{\rm{eff}}$ when phase and gain errors exist) through training. This can be formulated as a CS problem and the OMP can be applied to solve it \cite{channel estimation 1}, \cite{channel estimation via OMP}, especially for channels with a small number of paths, i.e.,  
$L_p \ll \min(N_r, N_t)$. 
Ignoring the angle quantization error and using the virtual channel representation, $\mb H$ is modeled as \cite{switches or phase shifters}, \cite{compressed channel sensing}, \cite{virtual model},
\begin{equation}
\label{CS model}
\mb H =\mb A_{r}\mb H_v\mb A_{t}^H, 
\end{equation}
where $\mb A_{r}\in \mathbb{C}^{N_r\times{G_r}}$ and $\mb A_{t}\in \mathbb{C}^{N_t\times{G_t}}$ are two dictionary matrices, and $\mb H_v\in\mathbb{C}^{G_r\times G_t}$ is a sparse matrix that contains the path gains of the quantized directions.  The two dictionary matrices  $\mb A_{r}$ and $\mb A_{t}$  are commonly constructed using array response vectors \cite{switches or phase shifters}. Vectorizing (\ref{CS model}) leads to 
\begin{equation} {\rm vec}(\mb H)=\bm \Psi \mb x,\end{equation}  
where  
\begin{equation}
\label{psi}
\mb \Psi =  \mb A_{t}^\ast \otimes \mb A_{r} 
\end{equation} 
is the basis matrix, $(\cdot)^\ast$ denotes the conjugate, and  \[\mb x \triangleq \mathrm{vec}(\mb H_v)\] is an $L_p$-sparse vector. Noisy observations of linear combinations of the entries of ${\rm vec}(\mb H)$  may be obtained by training, yielding 
\begin{equation}
\label{yvector} 
{\mb y = \mb \Phi {\rm vec}(\mb H) + \mb z =  \mb \Phi \mb \Psi \mb x + \mb z,}
\end{equation}  
where $\mb \Phi$ is the sensing matrix specified by the training scheme and $\mb z$ is the noise. 
The OMP method finds $L_p$ out of $G_rG_t$ candidate direction pairs in the dictionary, where $G_r$ and $G_t$ are the numbers of grid points for the AoA and AoD, respectively. The two dictionary matrices $\mb A_r$ and $\mb A_t$ can be designed to be unitary matrices when $G_t=N_t$ and $G_r=N_r$, and are redundant when $G_t>N_t$ and $G_r>N_r$. The computational complexity of the OMP method is about $O(NL_pG_tG_r)$, where $N$ is the number of observations. In general, the larger the number of grid points the better the performance, yet the heavier the computational burden and storage space. 

The above CS scheme assumes the array response vector is known so that the channel can be modeled as (\ref{CS model}), which is sparse on the basis built as (\ref{psi}). However, when phase errors and gain errors exist, it is the effective channel $\mb H_{\rm{eff}}$ rather than $\mb H$ to be estimated. The basis for $\mb H_{\rm{eff}}$ is hard to construct due to the unknown ${\mb E}_r$ and $\mb E_t$, and thus leading to a basis mismatch issue \cite{basis mismatch}, which may cause significant performance degradation of the CS estimators that rely on the basis.
 In the following, we propose an MC-based channel estimation scheme compatible with the hybrid system and does not rely on the basis; thus, it is effective for systems having arrays with phase errors and gain errors. 


\section{MC-Based Channel Estimation}
In this section, we first introduce the MC formulation of the channel estimation problem and design a training scheme that is compatible with the hybrid system. A GCG-Alt estimator is then proposed to solve the channel estimation problem. We finally generalize our approach to an IMC scheme.

\subsection{MC Formulation}
We propose to formulate the channel estimation problem as an MC problem including estimating a subset of the entries of $\mb H$ and recovering the full channel matrix by exploiting the low-rank nature of the channel and MC techniques. Define a sampling operator $P_{\Omega}(\cdot)$ as 
\begin{equation}  
\label{operator}
[P_{\Omega}( \mb H)]_{i,j}=
\begin{cases}
[ \mb H]_{i,j},    & \quad (i,j) \in \Omega\\
0,	   & \quad  \text{otherwise}
\end{cases}, 
\end{equation} 
where $[\mb H]_{i,j}$ denotes the $(i, j)$-th entry of $\mb H$ and $\Omega$ represents the sampling domain.
Let $p$ be the sampling density, then the number of sampled entries of $\mb H$ in the operator $P_{\Omega}(\cdot)$ is $N=pN_tN_r$. As suggested in \cite{exact MC}, $p\geq C{\widetilde{n}}^{1.2}r_{\rm{ch}}{\rm{log}}(\widetilde{n})/(N_rN_t)$ to guarantee recovery, where $\widetilde{n}=\text{max}(N_t,N_r)$, $C$ is a positive 
constant independent of $(\widetilde{n}, r_{\rm{ch}}, p$) and can be different for different types of matrices. For the mmWave channel, examples of $p=0.14$ and $p=0.5$ are seen \cite{MC_CS}, \cite{WCSP}.
In the noisy scenario, we obtain $P_{\Omega}(\mb H_{\rm{N}})$, where $\mb H_{\rm{N}}=\mb H+\mb E$ and $\mb E$ is the noise matrix. Then the full channel matrix is recovered by solving the low-rank recovery problem \cite{matrix completion with noise}
\begin{equation} 
\label{nyMC}
\min_{\widehat{\mb H}} {\rm rank}(\widehat{\mb H}), \quad \quad \mathrm{s.t.} \quad \|P_{\Omega}( \widehat{\mb H}- \mb H_{\rm{N}})\|^2_F\leq \delta^2.
\end{equation}
If $\mb E$ is white Gaussian noise with standard deviation $\sigma$, then $\|P_{\Omega}(\mb E)\|^2_F\leq (N+\sqrt{8N})\sigma^2$ with high probability \cite{matrix completion with noise} and $N=pN_rN_t$ is the total number of observations.
In the formulation of (\ref{nyMC}), the objective is to find an $\widehat{\mb H}$ with the minimum rank based on the noisy observations. There are no assumptions on the array responses. This is different from the CS-formulation in which the channel is represented as (\ref{CS model}) that relies on the array response for constructing $\mb A_r$ and $\mb A_t$.

The above MC problem is NP-hard and usually solved by using approximate algorithms. The singular value thresholding  (SVT) algorithm in \cite{SVT} and the fixed point continuation (FPC) algorithm in \cite{FPC} tackle this problem by using matrix shrinkage. 
They require full singular value decomposition (SVD) calculation at each iteration, which can yield high computational complexity when the size of $\mb H$ is large. The singular value projection (SVP) algorithm \cite{SVP}, \cite{Joint CSIT MC} solves the MC problem based on the classical projected gradient algorithm; the alternating minimization algorithm \cite{MCAltMin} converts the target matrix into its bi-linear form, i.e., $\mb H=\mb U\mb V^H$, and solves $\mb U$ and $\mb V$ alternatively. They both have lower computational complexity compared to SVT and FPC, but need to know the channel rank $r_{\rm ch}$, which is unknown in practice. Note that knowing $r_{\rm ch}$ can also help reduce the computational complexity of SVT and FPC as rank-$r_{\rm ch}$ SVD can be used instead. 
In this paper, we adopt a generalized conditional gradient (GCG) framework \cite{GCG} to reconstruct $\mb H$, which does not require $r_{\rm{ch}}$ and has lower computational complexity compared to SVT and FPC.

\subsection{Training Process}
\label{training} 
The sampling pattern specified by the sampling operator $P_{\Omega}(\cdot)$ has a crucial influence on the performance of MC algorithms. 
From \cite{matrix completion with noise}, at least one entry must be sampled from each row and each column to recover the original matrix. 
In this paper, we adopt the uniform spatial sampling (USS) scheme \cite{array signal MC}, which is proposed for array signal processing and outperforms alternative sampling schemes such as the Bernoulli scheme \cite[Section IV]{matrix completion with noise}. 
Following the USS sampling scheme, we take $N/N_t$ distinct noisy samples from the $N_r$ entries of each column of the channel matrix. 
During training, suppose one symbol is transmitted at each training stage and employ $M$ training stages with $S$ training steps at each training stage. At the BS, a unique processing vector $\mb f$ of (\ref{received}) is used at the $m$-th training stage, which will be denoted by $\mb f_m$. At the $m$-th stage, $\mb f_m$ remains unchanged and the MS changes the receiving processing matrix $\mb W$ by $S$ times. In the following, we use $\mb W_{m,s}$ to represent the MS receiving processing matrix at the $s$-th step of the $m$-th stage. 

The total number of training steps is $MS$. At the $s$-th step of the $m$-th training stage, the BS sends out one symbol $s_{m,s}$ with power $P$ through $\mb f_m$ and the MS receives $N_{m,s}\leq K_r$\footnote{The MS with only $K_r$ RF chains can only produce up to $K_r$ estimates simultaneously.} signals through $\mb W_{m,s}$. In this way, the observation at the $s$-th step of the $m$-th stage is 
\begin{equation}
\label{y_ts}
\mb y_{m,s}=\mb W_{m,s}^H\mb H\mb f_m s_{m,s}+\mb W_{m,s}^H\mb n_{m,s},
\end{equation}
where 
$\mb n_{m,s}\in\mathbb{C}^{N_r}$ is the noise vector. Assume all transmitted symbols during the training are identical and $s_{m,s}=\sqrt{P}$. By setting $\|\mb f_m\|^2_F=1$, the total transmitting power is $\|\mb f_m s_{m,s}\|_F^2=P$.
We define 
the pilot-to-noise ratio $\rm{(PNR)}$ as 
\begin{equation} 
\label{PNR} 
{\rm{PNR}}=\frac{\|\mb f_m s_{m,s} \|^2_F}{\sigma^2}, 
\end{equation}  
where the noise is assumed to be an additive white Gaussian noise (AWGN) with variance $\sigma^2$.

Collect all the $S$ received vectors of the $m$-th training stage into vector
\begin{equation}
\label{y_t}
\mb y_m =\sqrt{P}\mb W_m^H\mb H\mb f_m+\mb n_m,
\end{equation}
\vspace{-1ex}
where
\[\mb y_m=[\mb y^T_{m,1}, \mb y^T_{m,2}, \ldots, \mb y^T_{m,S}]^T,\]
\[ \mb W_m=[\mb W_{m,1}, \mb W_{m,2}, \ldots,\mb W_{m,S}],\]
\[\mb n_m=[\mb n^T_{m,1}\mb W^{\ast}_{m,1}, \mb n^T_{m,2}\mb W^{\ast}_{m,2}, \ldots,\mb n^T_{m,S}\mb W^{\ast}_{m,S}]^T,\]
and $(\cdot)^T$ represents the transpose.
Stacking all the received vectors from the $M$ training stages into matrix $\mb Y$ yields
\begin{equation}
	\label{Y}
	\mb Y=\sqrt{P}\mb W^H\mb H\mb F+\mb N,
\end{equation}
where $\mb Y=[\mb y_1,\mb y_2,\ldots,\mb y_M]$, $\mb N=[\mb n_1,\mb n_2,\ldots,\mb n_M]$, $\mb W=[\mb W_1,\mb W_2,\ldots,\mb W_M]$ and $\mb F=[\mb f_1,\mb f_2,\ldots,\mb f_M]$.
In the hybrid system, $\mb W$ and $\mb F$ of (\ref{Y}) are composed of RF beamformers and digital processors. At the $m$-th stage, $\mb f_m=\mb G_m  \mb b_m$, where $\mb G_m\in\mathbb{C}^{N_t\times K_t}$ and $ \mb b_m\in\mathbb{C}^{K_t}$
are the RF beamformer and digital processor at the BS, respectively; 
for the MS, 
\[\mb W_m=[\mb Q_{m,1} \mb D_{m,1},\ldots, \mb Q_{m,S} \mb D_{m,S}]\]
where $\mb Q_{m,s} \in\mathbb{C}^{N_r\times K_r}$ and $\mb D_{m,s}\in\mathbb{C}^{K_r\times N_{m,s}}$ are the RF beamformer and digital processor, respectively. The constraint of analog phase shifters requires $[\mb Q_{m,s}]_{i,j}\in\mathcal{W}_{\rm{RF}}$ and $[\mb  G_m]_{i,j}\in\mathcal{F}_{\rm{RF}}$, where $\mathcal{W}_{\rm{RF}}$ and $\mathcal{F}_{\rm{RF}}$ are two sets that contain all the possible phase shifts $e^{j2\pi k/2^I},k=0,1,\ldots,2^I-1$ of the MS and BS phase shifters, respectively, 
where $I$ is the number of bits of the phase shifter.

We design $\mb f_m$ to sample one column of $\mb H$ at each stage and choose $\mb W_m$ to sample $N/N_t$ distinct entries of that column.  
We set $M\geq N_t$ to guarantee that every column in $\mb H$ is sampled at least once. Let $j_m={\rm mod}(m, N_t)+1$, where $\rm mod(\cdot)$ denotes the modulus operation. 
At the $m$-th stage, 
\begin{equation} 
\label{fmexp}
\mb f_m\triangleq[0, \ldots, 1, \ldots, 0]^T
\end{equation}
 is set with 1 at its $j_m$-th entry, such that the $j_m$-th column of $\mb H$ is extracted. Since $\mb f_m=\mb G_m \mb b_m$, the design task is as follows:
\begin{align}
\label{design task F}\nonumber
\text{Find }&\mb G_m, \mb b_m, \\ \nonumber
\text{s.t. } &\mb G_m\mb b_m=\mb f_m,\\  
&\mb G_m\in\mathcal{F}_{\rm{RF}}^{N_t \times K_t}.
\end{align}
In order to satisfy the constraint of $\mb f_m$, the inner product of the $j_m$-th row of $\mb G_m$ and $\mb b_m$ must be $1$ and the other $N_t-1$ rows in $\mb G_m$ must be orthogonal to $\mb b_m$.

 We first present the design of $\mb G_m$ and $\mb b_m$ with $\mb f_m=[1,0,\ldots,0]^T$. Write $\mb G_m$ as 
\begin{equation}
\label{splitRF}
\mb G_m=\begin{bmatrix}
\mb G_{1,m}\\
 \mb G_{2,m}
\end{bmatrix}, 
\end{equation}
where $\mb G_{1,m}\in\mathcal{F}_{\rm{RF}}^{2\times K_t},   \mb G_{2,m} \in\mathcal{F}_{\rm{RF}}^{(N_t-2)\times K_t}$,
and then $\mb G_m\mb b_m=\mb f_m$ in (\ref{design task F}) splits into 
\begin{align}
\mb G_{1,m}\mb b_m&=\mb e_1 \label{cond1F},\\
\mb G_{2,m}\mb b_m&=\mb 0_{(N_t-2)\times 1} \label{cond2F},
\end{align}
where $\mb e_1=[1,0]^T$. Since the entries in $\mb G_{1,m}$ cannot be $0$, we need $K_t\geq 2$ to guarantee that problem (\ref{cond1F}) is solvable. This is because if $K_t=1$, the vector $\mb b_m$ becomes a scalar $b_m$. Then problem (\ref{cond1F}) becomes
\begin{equation}
\label{ana}
\mb G_{1,m}b_{m}=\begin{bmatrix}
                                     1\\
                                     0
                                 \end{bmatrix},
\end{equation}
which has no solution unless the entries in $\mb G_{1,m}$ can be $0$.

If $\mb G_{1,m}$ is known, the least square solution of (\ref{cond1F}) is
\begin{equation}
\label{fbbsol}
\mb b_m = \mb G_{1,m}^H (\mb G_{1,m} \mb G_{1,m}^H)^{-1}\mb e_1.
\end{equation}
We can see that $ \mb G_{1,m}\mb G_{1,m}^H$ should be invertible, which requires $\mb G_{1,m}$ having full row rank.
Considering $[\mb G_{1,m}]_{i,j}\in\mathcal{F}_{\rm{RF}}^{2\times K_t}$, the Vandermonde matrix is a natural choice for $\mb G_{1,m}$. Therefore, we construct $\mb G_{1,m}$ as
\begin{align}
\label{FRF1_con}\nonumber
&[\mb G_{1,m}]_{1,l}=\frac{1}{\sqrt{K_t}}\omega^{l-1}_1, \quad [\mb G_{1,m}]_{2,l}=\frac{1}{\sqrt{K_t}}\omega^{l-1}_2,\\\nonumber
&\quad\quad\quad \quad \quad \quad \quad l=1, 2,\ldots, K_t,
\end{align}
where
\begin{equation}
\label{whereCond}\nonumber
 \omega_1=e^{jn_1\frac{2\pi}{2^I}}, \quad  \omega_2=e^{jn_2\frac{2\pi}{2^I}}, 
\end{equation}
and $n_1, n_2$ are integers\footnote{For achieving high numerical stability, we can choose proper $\omega_1$ and $\omega_2$ so that $\mb G_{1,m}\mb G^H_{1,m}$ is well-conditioned.}. Here we require $n_1\neq n_2$ so that $\mb G_{1,m}$ has full row rank. 
The minimum requirement for realizing $\mb f_m$ is $I=1, K_t=2$. For example, when $I=1, K_t=2$, $\mb G_{1,m}\in \mathcal{F}_{\rm{RF}}^{2\times 2} $, and (\ref{fbbsol}) becomes
\begin{equation}
\label{fbbre}
\mb b_m=\mb G^{-1}_{1,m}\mb e_1.
\end{equation}
With $I=1$, choosing $n_1=0, n_2=1$, we have $\mb G_{1,m}$ and $\mb b_m$ as
\begin{equation}
\label{Gb}
\mb G_{1,m}=\begin{bmatrix}
1&1\\
1&e^{j\pi}
\end{bmatrix},\quad\quad \mb b_m=\begin{bmatrix}
1/2\\
1/2\\
\end{bmatrix}.
\end{equation}

textcolor{blue}{After obtaining $\mb b_m$, all the rows of $\mb G_{2,m}$ in (\ref{cond2F}) can be chosen as $\mb G_{1,m}(2,:)$ since $\mb G_{1,m}(2,:)\mb b_m=0$. This produces  
\begin{equation}
\label{FRFtdesign}
\mb G_m=\begin{bmatrix}
\mb G_{1,m}\\
\mb G_{1,m}(2,:)\\
\vdots\\
\mb G_{1,m}(2,:)
\end{bmatrix}.
\end{equation}
For other $\mb f_m$ with the $j_m$-th entry being $1$, we only need to swap the first and the $j_m$-th row of the $\mb G_m$ in (\ref{FRFtdesign}) and keep the designed $\mb b_m$ unchanged\footnote{The approach in \cite{RFbaseband} also solves problem (\ref{design task F}), and its solution is equivalent to our solution when $I=1, K_t=2$.}.

During each of the $S$ training steps, the MS produces the estimates of $N_{m,s}$ entries of the $j_m$-th column of $\mb H$ through $\mb W_{m,s}\in\mathbb{C}^{N_r\times N_{m,s}}$. 
Let $i_{m,s,q}$ be the row index of the $q$-th sampled entry and 
\[ \mathcal I_{m,s}=\{i_{m,s,1}, i_{m,s,2},\ldots,i_{m,s,N_{m,s}}\}.\] 
In order to achieve interference-free sampling, the required $\mb W_{m,s}$ is constructed as 
\begin{equation}  
\label{operatorII}
[\mb W_{m,s}]_{i,j}=
\begin{cases}
1,    & \quad i = i_{m,s,j}, j=1,2,\ldots, N_{m,s }\\
0,	   & \quad  \text{otherwise}
\end{cases}. 
\end{equation}
Then the design task is as follows:
 \begin{align}
 \label{design task W}\nonumber
 \text{Find }&\mb Q_{m,s}, \mb D_{m,s}, \\ \nonumber 
 \text{s.t. }& \mb Q_{m,s}\mb D_{m,s}=\mb W_{m,s}\\  
 &\mb Q_{m,s}\in\mathcal{W}_{\rm{RF}}^{N_r \times K_r}.
 \end{align}
We first present the design of $\mb Q_{m,s}$ and $ \mb D_{m,s}$ with $\mathcal I_{m,s}=\{1, 2, \ldots,N_{m,s}\}$, which means
\begin{equation}
\label{Wts}
\mb W_{m,s}=\begin{bmatrix}
\mb I_{N_{m,s}}\\
\mb 0 
\end{bmatrix}.
\end{equation}
Write $\mb Q_{m,s}$ as
\begin{equation}
\label{WRFts}
\mb Q_{m,s}=\begin{bmatrix}
\mb Q_{1,m,s}\\
\mb Q_{2,m,s}
\end{bmatrix},
\end{equation}
where $\mb Q_{1,m,s}\in\mathbb{C}^{K_r\times K_r}$ and $\mb Q_{2,m,s} \in\mathbb{C}^{(N_r-K_r) \times K_r}$. Then $\mb Q_{m,s} \mb D_{m,s} =\mb W_{m,s}$ in (\ref{design task W}) splits into
\begin{align}
\mb Q_{1,m,s} \mb D_{m,s}&=\mb W_{1,m,s},\label{cond1W}\\
\mb Q_{2,m,s} \mb D_{m,s}&=\mb 0 
\label{cond2W},
\end{align}
where $\mb D_{m,s}\in\mathbb{C}^{K_r\times N_{m,s}}$, and 
\begin{equation}
\mb W_{1,m,s}=\begin{bmatrix}
\mb I_{N_{m,s}}\\
\mb 0_{(K_r-N_{m,s})\times N_{m,s}}
\end{bmatrix}. 
\end{equation}
Note that we need $N_{m,s}\leq K_r-1$ to guarantee only one $1$ in each column of $\mb W_{m,s}$. If $\mb Q_{1,m,s}$ is given, the solution of (\ref{cond1W}) is
\begin{equation}
\label{WBBsol}
\mb D_{m,s}= \mb Q_{1,m,s}^{-1}\mb W_{1,m,s}.
\end{equation}
$\mb Q_{1,m,s}$ should be invertible. 
Similar to the design of $\mb G_{1,m}$,
we construct $\mb Q_{1,m,s}$ as 
\begin{equation}
\label{WRF1design}
[ \mb Q_{1,m,s}]_{k,l}=\frac{1}{\sqrt{K_r}}\omega^{l-1}_k, k,l=1, 2, \ldots, K_r,
\end{equation}
where 
\begin{equation}
\omega_k=e^{jn_k\frac{2\pi}{2^I}}, 
\end{equation}
After obtaining $\mb D_{m,s}$, all the rows of $\mb Q_{2,m,s}$ in (\ref{cond2W}) can be chosen as $\mb Q_{1,m,s}(K_r,:)$ as $\mb Q_{1,m,s}(K_r,:)\mb D_{m,s}=\mb 0_{1\times N_{m,s}}$, yielding  
\begin{equation}
\label{WRFtsW}
\mb Q_{m,s}=\begin{bmatrix}
\mb Q_{1,m,s}\\
\mb Q_{1,m,s}(K_r,:)\\
\vdots\\
\mb Q_{1,m,s}(K_r,:)
\end{bmatrix}.
\end{equation}
For other $\mathcal I_{m,s}$, we just need to permute the rows in the $\mb Q_{m,s}$ of (\ref{WRFtsW}) according to the elements in $\mathcal I_{m,s}$ and keep the designed $\mb D_{m,s}$ unchanged. For example, if $\mathcal I_{m,s}=\{1, N_r, 3,4, \ldots, N_{m,s}\}$, we swap the second and the $N_r$-th row of $\mb Q_{m,s}$. Note that when $K_r=2$, we have $N_{m,s}=K_r-1=1$ and $\mb D_{m,s}$ becomes a vector $\mb d_{m,s}$, so that problem (\ref{design task W}) reduces to problem (\ref{design task F}). Similarly, we require $I\geq 1$ and $K_r\geq 2$ for realizing $\mb W_{m,s}$.


The processing matrices designed above are applied to (\ref{Y}) to obtain the received samples in $\mb Y$. Without loss of generality, let the transmitted symbol power $P=1$. We can then 
construct a matrix $\widetilde{\mb H} \in\mathbb{C}^{N_r\times N_t}$ using $\mb Y \in\mathbb{C}^{N/M\times M}$ as
\begin{equation}
\label{H_tilde}
 [\widetilde{\mb H}]_{l,k}=
\begin{cases}
[\mb Y]_{i_{l,k},j_{l,k}}, &(l,k)\in\Omega,\\
&  i_{l,k}=1,\ldots,\frac{N}{M}, j_{l,k}=1,\ldots,M \\
0,             & \text{otherwise,}
\end{cases}
\end{equation}
where $\Omega$ contains the positions of all $N$ samples stored in the form of $(l, k)$ with $l\in[1,N_r]$ and $k\in[1,N_t]$ and $(l,k)$ indicates sampling the $(l,k)$-th entry of $\widetilde{\mb H}$. In the above, $(i_{l,k},j_{l,k})$ represents the index of the corresponding entry in $\mb Y$ for the $(l,k)$-th entry of $\widetilde{\mb H}$. 
Note that $\widetilde{\mb H}$ and $P_{\Omega}(\widetilde{\mb H})$ are actually the same.
Then the channel matrix $\mb H$ can be estimated from $ P_{\Omega}(\widetilde{\mb H})$ by using MC algorithms.

%

\emph{Remark 1}: The proposed training scheme can also be applied to switch-based hybrid systems \cite{WCSP}, as the processing matrices $\mb F$ and $\mb W$ that contain only 1's and 0's can be directly achieved by switching on and off the switches. When the array-inherent impairments are present, the samples obtained by the proposed training scheme are noisy observations of the entries of the effective channel matrix $\mb H_{\rm{eff}}=\mb E_r\mb H\mb E^H_t$. Therefore, the MC-based estimator estimates $\mb H_{\rm{eff}}$ instead of $\mb H$.


\subsection{GCG-Alt Estimator}
In this paper, we adopt the framework introduced in \cite{GCG} that consists of a relaxed GCG algorithm and a local search algorithm to estimate $\mb H$ (or $\mb H_{\rm{eff}}$). We propose an alternating minimization (AltMin) algorithm as the local search algorithm and thus name the resulting estimator as the GCG-Alt estimator. 
This estimator utilizes the relaxed GCG algorithm to generate a good initial estimate, based on which the AltMin algorithm converges fast to an optimized solution. 

The problem is formulated as
\begin{equation}
\label{main_eq}
\min_{\widehat{\mb H}\in\mathbb{C}^{N_r\times N_t}} \quad \phi(\widehat{\mb H}),
\end{equation}
where 
\[\phi(\widehat{\mb H})\triangleq
 f(\widehat{\mb H})+\mu\|\widehat{\mb H}\|_{\ast},
\]
\begin{equation}
\label{feq}\nonumber
f(\widehat{\mb H}) \triangleq\frac{1}{2}\|P_{\Omega}(\widehat{\mb H})-P_{\Omega}(\widetilde{\mb H} )\|_F^2,
\end{equation}
 $\mu>0$  is a regularization coefficient and $\|\widehat{\mb H}\|_\ast$ is the nuclear norm (i.e., summation of the singular values) of $\widehat{\mb H}$. 

\subsubsection{Relaxed GCG Algorithm}
Following \cite{GCG}, problem (\ref{main_eq}) can be solved via the GCG algorithm by successively finding the descent direction $\mb Z$ of $f(\widehat{\mb H})$ and updating $\widehat{\mb H}$ by $(1-\eta)\widehat{\mb H}+\theta\mb Z$, where $\eta\in [0,1]$ is the step size properly chosen to avoid divergence and $\theta$ is a parameter chosen to minimize $\phi(\widehat{\mb H})$. At the $k$-th iteration, $\mb Z_{k}$ is found as \cite{GCG}:
\begin{align}
\label{GCG solution}
&\mb Z_{k}=\min_{\|\mb Z\|_\ast\leq 1} \langle\mb Z, \nabla f(\widehat{\mb H}_{k-1})  \rangle,
\end{align}
where $\nabla$ represents the gradient and \[ \langle\mb A,\mb B\rangle \triangleq \text{tr}(\mb A^H\mb B)\] represents the inner product of two matrices. The solution to (\ref{GCG solution}) is given \cite{GCG} as
\begin{equation}
\label{directionZ}
\mb Z_{k}={\mb u}_{k-1}{\mb v}^H_{k-1},
\end{equation}
where $({\mb u}_{k-1}, {\mb v}_{k-1})$ is the top singular vector pair of 
\begin{equation}
\label{gradient}
-\nabla f(\widehat{\mb H}_{k-1})=-P_{\Omega}(\widehat{\mb H}_{k-1}-\widetilde{\mb H}).
\end{equation}
Then we have 
\begin{equation}
\label{update}
\widehat{\mb H}_{k}=(1-\eta_k)\widehat{\mb H}_{k-1}+\theta_k\mb Z_{k}.
\end{equation}
Following \cite{GCG}, $\theta_k$ can be chosen as 
\begin{equation}
\label{theta_k1}
\widetilde{\theta}_k=\text{arg}\min_{\theta_{k}\geq 0}\phi (\theta_k),
\end{equation}
 where 
\begin{equation}
\label{updatePhi}
\phi (\theta_k)\triangleq f((1-\eta_k)\widehat{\mb H}_{k-1}+\theta_k\mb Z_k)+\mu\|(1-\eta_k)\widehat{\mb H}_{k-1}+\theta_k\mb Z_k\|_\ast.
\end{equation}

However, solving (\ref{theta_k1}) can be computational expensive since it involves the evaluation of  $\|(1-\eta_k)\widehat{\mb H}_k+\theta_k\mb Z_k\|_\ast$. In order to reduce the computational complexity, \cite{GCG} proposes to minimize an upper bound of $\phi(\theta_k)$, which is 
\begin{align}
\label{upper_bound} 
h(\theta_k)&=f((1-\eta_k)\widehat{\mb H}_{k-1}+\theta_k\mb Z_k)+\mu(1-\eta_k)\|\widehat{\mb H}_{k-1}\|_\ast+\mu\theta_k. 
\end{align}
This upper bound is obtained by using the convex property of the nuclear norm that 
\[ \|(1-\eta_k)\widehat{\mb H}_{k-1}+\theta_k\mb Z_k\|_\ast\leq (1-\eta_k)\|\widehat{\mb H}_{k-1}\|_\ast+\theta_k\|\mb Z_{k}\|_\ast \]  and the fact that \[ \|\mb Z_{k}\|_{\ast}\leq 1.\]
Then we have
\begin{equation}
\label{theta_k2}
\widetilde{\theta}_k=\text{arg}\min_{\theta \geq 0}h(\theta_k).
\end{equation}
The solution of (\ref{theta_k2}) is obtained by letting \[ \partial h(\theta_k)/\partial \theta_k=0\] as
\begin{subequations}\label{theta_kso}
\begin{equation}
\theta_k=\frac{2\mathcal{R}({\mb z_{k\Omega}^H} \widetilde{\mb h}_\Omega)-(1-\eta_k){\mb z^H_{k\Omega}}\widehat{\mb h}_{k\Omega}-2\mu}{2{\mb z^H_{k\Omega}}\mb z_{k\Omega}},
\tag{\ref{theta_kso}}
\end{equation}
where 
\begin{align}
           \mb z_{k\Omega}&= {\rm vec}(P_{\Omega}(\mb Z_k)) \label{z_rep},\\
\widetilde{\mb h}_\Omega={\rm vec}(P_\Omega( &\widetilde{\mb H} )), \quad \widehat{\mb h}_{k\Omega} =  {\rm vec}(P_{\Omega}(\widehat{\mb H}_{k-1}))\label{h_rep},
\end{align}
\end{subequations}
where $\rm{vec}(\cdot)$ denotes vectorization and $\mathcal{R}(\cdot)$ denotes the real part of a number.
After obtaining $\theta_k$, we can update $\widehat{\mb H}_{k}$ as in (\ref{update}).
The $k$-th iteration of the GCG algorithm updates $\widehat{\mb H}$ from $\widehat{\mb H}_{k-1}$ to $\widehat{\mb H}_k$ by adding a rank-1 matrix $\theta_k\mb Z_{{k}}$, producing an estimate $\widehat{\mb H}_{k}$ of rank $k$.

\subsubsection{AltMin Algorithm}
Recall that the parameter $\theta_k$ is chosen based on an upper bound of the objective function in (\ref{updatePhi}). This suggests that $\widehat{\mb H}_{k}$ in (\ref{update}) may not be the optimal solution and it is possible to obtain a solution with rank $k$ that improves $\widehat{\mb H}_{k}$. Therefore, 
before moving to the next iteration of the relaxed GCG algorithm, a local search algorithm can be applied to find such a solution using $\widehat{\mb H}_{k}$ as the input and compute iteratively an output $\widehat{\mb H}^{Q}_{k}$ with rank $k$ and \[\phi(\widehat{\mb H}^{Q}_{k})<\phi(\widehat{\mb H}_{k}),\]
where $Q$ is the number of iterations of the local search algorithm. Following \cite{GCG}, the nuclear norm  of $\widehat{\mb H}$ can be written as
\begin{equation}
\label{nunorm}
\|\widehat{\mb H}\|_{\ast}=\frac{1}{2}\min_{{\mb U},{\mb V}}\{\|{\mb U}\|_F^2+\|{{\mb V}} \|_F^2: \widehat{\mb H}={\mb U}{{\mb V}}^H\},
\end{equation}
where $ {\mb U}\in\mathbb{C}^{N_r\times \widehat{r}}$ and $ {\mb V}\in\mathbb{C}^{N_t\times \widehat{r}}$ with $\widehat{r}$ being the rank of $\widehat{\mb H}$. Therefore, finding an $\widehat{\mb H}$ to minimize the objective function in (\ref{main_eq}) becomes finding a pair of $(\mb U, \mb V)$ to minimize
\begin{equation}
\label{main_eq2} 
\widetilde{\phi}(  \mb U, {\mb V})\triangleq f( {\mb U}{ {\mb V}}^H)+\frac{1}{2}\mu(\| {\mb U}\|_F^2+\|{ {\mb V}} \|_F^2).
\end{equation}
Given \[\widehat{\mb H}_{k-1}= {\mb U}_{k-1} {\mb V}^H_{k-1},\] the update $\widehat{\mb H}_{k}$ in (\ref{update}) obtained by the relaxed GCG algorithm is equivalent to the following:
\begin{align}
\label{UV_k}\nonumber
 {\mb U}_{k}&=[\sqrt{1-\eta_k} {\mb U}_{k-1},\sqrt{\theta_{k}} {\mb u}_{k-1}],\\
 {\mb V}_{k}&= [\sqrt{1-\eta_k} {\mb V}_{k-1},\sqrt{\theta_{k}} {\mb v}_{k-1}],
\end{align}
where \[\mb Z_{k}= {\mb u}_{k-1} {\mb v}^H_{k-1},\] $ {\mb U}_{k}\in\mathbb{C}^{N_r \times {k}}$ and $ {\mb V}_{{k}}\in\mathbb{C}^{N_t\times {k}}$.

Let us use the $k$-th update of $( \mb U_k,  \mb V_k)$ obtained by the relaxed GCG algorithm as the input of the AltMin algorithm. We now discuss the update of $\mb V_k^i$ at the $i$-th update of the AltMin algorithm. Define
\begin{equation}
\label{V}
\widetilde{\phi}(  \mb V | \mb U ^{i-1}_{k})=\frac{1}{2}\|P_{\Omega}(\widetilde{\mb H}- {\mb U}^{i-1}_{k}  {\mb V}^H)\|^2_F+\frac{\mu}{2}\| {\mb V}\|^2_F. 
\end{equation}
Vectorizing ${\mb V}$ in (\ref{V}) into $ {\mb v}$, we have
\begin{align}
\nonumber
\widetilde{\phi}( {\mb v}| {\mb U}^{i-1}_{k})&=\frac{1}{2}\|\widetilde{\mb h}_{\Omega}-P_{\Omega}((\mb I_{N_t}\otimes {\mb U}^{i-1}_{k}) {\mb v})\|^2_F+\frac{\mu}{2}\| {\mb v}\|^2_F\\
&=\frac{1}{2}\|\widetilde{\mb h}_{\Omega}- \bm{\mathcal{U}}^{i-1}_{k}  {\mb v}\|^2_F+\frac{\mu}{2}\| {\mb v}\|^2_F\label{v},
\end{align}
where 
\begin{align}
\label{cond12}
 \bm{\mathcal{U}}^{i-1}_{k} &=P_{\widetilde{\Omega}}\left(\mb I_{N_t}\otimes {\mb U}^{i-1}_{k} \right)\in \mathbb{C}^{N_tN_r\times N_tk},  
\end{align}
 $\widetilde{\Omega}$ stores the positions of the $N$ sampled entires out of the $N_t N_r$ entries of ${\rm{vec}}(\widetilde{\mb H})$ and the operator $P_{\widetilde{\Omega}}(\mb A)$ keeps the rows of $\mb A$ corresponding to $\widetilde{\Omega}$ while sets other rows of $\mb A$ to zero. 
Given $\mb U_k^{i-1}$, \[ {\mb V}^{i}_{k} ={\rm vec}^{-1}( {\mb v}^{i}_{k})\] can be updated by solving 
\begin{equation}
\label{V_f}
 {\mb v}^{i}_{k}=\min_{ {\mb v}}\widetilde{\phi}\left( {\mb v}| {\mb U}^{i-1}_{k}\right). 
\end{equation}
Since (\ref{v}) is a quadratic smooth function, the solution of (\ref{V_f}) can be found by solving
\[ \frac{\partial \widetilde{\phi}\left( {\mb v}|  {\mb U}^{i-1}_{k}\right)} { \partial  {\mb v}}=\mb 0. \]  
Therefore, we update ${\mb V}^{i}_{k}$ as
\begin{equation}
\label{uvform_sol}
 {\mb V}^i_{k}={\rm{vec}}^{-1}\left({\mb v}^i_{k}\right),
\end{equation}
where
\begin{equation}
\label{vform}
 {\mb v}^{i}_{k}=\left((\bm{\mathcal{U}}^{i-1}_{k})^H \bm{\mathcal{U}}^{i-1}_{k} +\mu \mb I_{N_tk}\right)^{-1}( \bm{\mathcal{U}}^{i-1}_{k} )^H\widetilde{\mb h}_{\Omega}.
\end{equation}
Following similar procedures, given $\mb V^i_k$, we can define  
\begin{align}
\label{cond12}
 \bm{\mathcal{V}}^{i}_k & =P_{\widetilde{\Omega}}\left(( {\mb V}^{i}_{k})^\ast\otimes \mb I_{N_r} \right) \in\mathbb{C}^{N_tN_r\times N_rk},
\end{align}
and update  
\begin{equation}
\label{uvform_sol}
 {\mb U}^{i}_{k}={\rm{ vec}}^{-1}(  \mb u ^i_{k}),\quad \end{equation}
where
\begin{equation}
\label{uform}
 {\mb u}^{i}_{k}=\left( (\bm{\mathcal{V}}^i_k)^H \bm{\mathcal{V}}^i_k + \mu\mb I_{N_rk}\right)^{-1}(\bm{\mathcal{V}}^i_k)^H\widetilde{\mb h}_{\Omega}. 
\end{equation}
The updates in (\ref{vform}) and (\ref{uform}) can be done iteratively for a number of iterations.  

\subsubsection{Stopping Criteria}
 Define the relative contribution of the $i$-th iteration of the AltMin algorithm as
\begin{equation}
\label{epsilon_p}
\epsilon^{i}_{k}=\frac{\widetilde{\phi}( {\mb U}^{i-1}_{k}, {\mb V}^{i-1}_{k})-\widetilde{\phi}( {\mb U}^{i}_{k}, {\mb V}^{i}_{k})}{\widetilde{\phi}( {\mb U}^{i-1}_{k}, {\mb V}^{i-1}_{k})} 
\end{equation}
 and a threshold $\epsilon_{a}$.
Then we stop the AltMin algorithm when $\epsilon^{i}_{k}\leq\epsilon_{a}$. Suppose the AltMin algorithm stops after $Q$ iterations, the output of the AltMin algorithm replaces the $k$-th update obtained by the relaxed GCG algorithm, i.e.,
\[( {\mb U}_{k}, {\mb V}_{k})\leftarrow( {\mb U}^{Q}_{k}, {\mb V}^{Q}_{k})\]

Similarly, we also set an energy threshold $\epsilon$ to determine whether the GCG-Alt estimator should stop iterating. Let the relative energy difference between the $k$-th and the $(k-1)$-th update of the GCG-Alt estimator be 
\begin{equation}
\label{epsilon_k+1}
\epsilon_{k}=\frac{\|\widehat{\mb H}_{k}\|^2_F-\|\widehat{\mb H}_{k-1}\|^2_F}{\|\widehat{\mb H}_{k-1}\|^2_F}.
\end{equation}
We can stop the estimator when $\epsilon_{k}\leq\epsilon$.  
In addition, by using our proposed training scheme, $P_{\Omega}(\widetilde{\mb H})$ is equivalent to $P_{\Omega}(\mb H+\mb N_h)$, where $\mb N_h\in\mathbb{C}^{N_r\times N_t}$ is the white Gaussian noise matrix. Assume the noise standard deviation is known as  $\sigma$, we have $\|P_{\Omega}(\mb N_h)\|^2_F\leq (N+\sqrt{8N})\sigma^2$ with large probability \cite{matrix completion with noise}. Define $\delta^2_k=\|P_{\Omega}(\widehat{\mb H}_k-\widetilde{\mb H})\|^2_F$, we introduce an additional stopping criterion that if
\begin{equation}
\label{st2}
\delta^2_k\leq (N+\sqrt{8N})\sigma^2,
\end{equation}
the estimator also stops.
The GCG-Alt estimator is summarized  in Algorithm 1.
\begin{table}
\begin{tabular}{ll}\rule{160mm}{.1pt}
\\
\textbf{Algorithm 1} \\ 
\rule[2mm]{160mm}{.1pt} 
\end{tabular}
\vspace{-2ex}
\begin{algorithmic}[1] 
\State \textbf{Input:} $P_{\Omega}(\widetilde{\bold H}), \mu,\epsilon, \epsilon_a$ 
\State \textbf{Initialization:} $ {\bold U}_0=\varnothing , {\bold V}_0=\varnothing,k=0,\epsilon_0=\infty $  
\While {$\epsilon_{k}>\epsilon$ or $\delta^2_k> (N+\sqrt{8N})\sigma^2$}
   \State $( {\bold u}_{k}, {\bold v}_{k})\gets$ top singular vector pair of $-\nabla f(\widehat{\bold H}_{k})$
    \State $k=k+1$
    \State $\eta_k\gets 2/(k+1)$， and determine $\theta_k$ using (\ref{theta_kso})
    \State $ {\bold U}_{k}\leftarrow [\sqrt{1-\eta_k} {\bold U}_{k-1},\sqrt{\theta_{k}} {\bold u}_{k-1}]$
    \State $ {\bold V}_{k}\leftarrow [\sqrt{1-\eta_k} {\bold V}_{k-1},\sqrt{\theta_{k}} {\bold v}_{k-1}]$
    \State \textbf{Initialization}:$i=0, \epsilon^{0}_{k}=\infty, ( {\bold U}^0_{k},  {\bold V}^0_{k})\leftarrow ( {\bold U}_{k}, {\bold V}_{k})$
               \While {$\epsilon^i_{k}>\epsilon_a$}
               \State $i=i+1$
                \State obtain $ {\bold U}^{i}_{k}$ and $ {\bold V}^{i}_{k}$ via (\ref{uform}) and (\ref{vform})
               \State calculate $\epsilon^{i}_{k}$ in (\ref{epsilon_p})
               \EndWhile 
    \State $( {\bold U}_{k}, {\bold V}_{k})\leftarrow ( {\bold U}^{i}_{k}, {\bold V}^{i}_{k})$
    \State calculate $\epsilon_{k}$ in (\ref{epsilon_k+1})
\EndWhile
\State \textbf{Output:} the estimated channel $\widehat{\bold H}=\widehat{\bold H}_{k}= {\bold U}_{k} {\bold V}^H_{k}$
\end{algorithmic}
\begin{tabular}{ll}
\rule[1.5mm]{160mm}{.1pt} 
\end{tabular}
\end{table}

\subsubsection{Computational Complexity}
Define a flop as an operation of real-valued numbers. We now analyze the computational complexity of the GCG-Alt estimator. For calculating the top singular vector pair in step 4 of Algorithm 1, the computational cost is $8(2q+3)(g+1)N_tN_r$ flops by using the Randomized SVD method in \cite{RandSVD}, where the exponent parameter $q=2$ and the oversampling parameter $g=10$. 
Calculating step 6 of Algorithm 1 requires $(4p+16)N_tN_r$ flops.
Suppose at the $k$-th iteration of the GCG algorithm, $\bm{\mathcal{U}}^{i-1}_{k}$ is a block diagonal matrix with each block of the size $N_r \times k$  and $N_t$ blocks in total, but there are only $pN_r\times k$ non-zero elements in each block. 
Therefore, the calculation of $\left((\bm{\mathcal{U}}^{i-1}_{k})^H \bm{\mathcal{U}}^{i-1}_{k} +\mu \mb I_{N_tk}\right)^{-1}$ only requires $8k^2pN_rN_t+4k^3N_t+8k^2N_t+kN_t\approx 8k^2pN_rN_t+4k^3N_t+8k^2N_t$ flops. 
The calculation of $( \bm{\mathcal{U}}^{i-1}_{k} )^H\widetilde{\mb h}_{\Omega}$ requires $8kpN_rN_t$ flops and the multiplication of $\left((\bm{\mathcal{U}}^{i-1}_{k})^H \bm{\mathcal{U}}^{i-1}_{k} +\mu \mb I_{N_tk}\right)^{-1}$ and $( \bm{\mathcal{U}}^{i-1}_{k} )^H\widetilde{\mb h}_{\Omega}$ requires $8k^2N_t$ flops. 
Therefore, the total number of flops needed for obtaining ${\mb V}^{i}_{k}$ is $(8k^2pN_r+4k^3+16k^2+8kpN_r)N_t$. Similarly, the total number of flops needed for obtaining ${\mb U}^{i}_{k}$ is $(8k^2pN_t+4k^3+16k^2+8kpN_t)N_r$. The calculations in step 13 and 16 of Algorithm 1 require way fewer flops than other steps in Algorithm 1 and are thus ignored. The flop counts are summarized in Table 1.

\begin{table*}  
\hskip-24mm
\small 
\begin{tabular}{l c c c}
\hline
\textbf{Table 1}  \\ 
\hline
\textbf{Algorithm} &\textbf{Operation}                     & \textbf{Flops per iteration}& \textbf{Total} \\
\hline
 \multirow{1}{*}{GCG} & Step 4 of Algorithm 1 & $8(2q+3)(g+1)N_rN_t$ & $8\widehat{r}_{\rm{GCG}}BN_tN_r$\\
($\widehat{r}_{\rm{GCG}}$ iterations) & Step 6 of Algorithm 1            & $(4p+16)N_rN_t$ &  where $B=(2q+3)(g+1)+(4p+16)$ \\ 
\hline
AltMin&\multirow{2}{*}{$\left((\bm{\mathcal{U}}^{i-1}_{k})^H \bm{\mathcal{U}}^{i-1}_{k} +\mu \mb I_{N_tk}\right)^{-1}( \bm{\mathcal{U}}^{i-1}_{k} )^H\widetilde{\mb h}_{\Omega}$}           & $8k^2pN_rN_t+4k^3N_t$      &\multirow{1}{*}{$\frac{1}{3}Q\widehat{r}_{\rm{GCG}}(\widehat{r}_{\rm{GCG}}+1)pN_rN_t(16\widehat{r}_{\rm{GCG}}+32)$}\\
($Q$ iterations)&  &  $+16k^2N_t+8kpN_rN_t$ & $+$  \\
$\bm{\mathcal{U}}^{i}_{k} \in\mathbb{C}^{N_tN_r \times kN_t}$&\multirow{2}{*}{$\left( (\bm{\mathcal{V}}^i_k)^H \bm{\mathcal{V}}^i_k + \mu\mb I_{N_rk}\right)^{-1}(\bm{\mathcal{V}}^i_k)^H\widetilde{\mb h}_{\Omega}$}           &       $8k^2pN_rN_t+4k^3N_r$   &{$\frac{1}{3}Q\widehat{r}_{\rm{GCG}}(\widehat{r}_{\rm{GCG}}+1)(N_t+N_r)(3\widehat{r}^2_{\rm{GCG}}+19\widehat{r}_{\rm{GCG}}+8)$}    \\
$\bm{\mathcal{V}}_k^i \in\mathbb{C}^{N_tN_r\times kN_r}$&     &       $+16k^2N_r+8kpN_rN_t$   \\
\hline
OMP&  & &\multirow{2}{*}{$8p\widehat{r}_{\rm{OMP}}N_tN_rG_tG_r$}\\
($\widehat{r}_{\rm{OMP}}$ iterations)\\
\hline
\end{tabular}
\end{table*}



 \subsection{Inductive Matrix Completion} 

\label{secIMC}

 In the training scheme proposed in Section \ref{training}, we essentially activate one transmitter antenna during each training stage and the total transmitted power $P$ is concentrated on a single transmitting antenna. 
This may be feasible in scenarios where the path loss of the transmission link is not significant, such as in the mmWave massive MIMO-based 
ultra-dense networks \cite{backhaul and access} where 
the path loss is even smaller than that in the conventional cellular networks \cite{backhaul and access}. 
For scenarios 
where the transmission distance is long and thus incurs a higher path loss, the peak transmission power for a single antenna can be high if a high $\rm{PNR}$ is required. 
In order to address this challenge, we propose to generalize the training scheme in Section \ref{training} following the principle of low-rank matrix recovery based on rank-1 measurements \cite{IMC}. 
With this generalization, all the transmitter antennas are activated simultaneously and the total transmitting power are spread out on the array, reducing the peak power transmitted from the antennas. The channel estimation problem is then reformulated as an inductive matrix completion (IMC) problem  \cite{IMC}, which can be solved directly by applying our proposed GCG-Alt estimator.

In the IMC framework, instead of directly sampling and completing $\mb H$, a transformed matrix 
\[ \mb C=\mb X_{\rm{L}}^H\mb H\mb X_{\rm{R}}\]
is first sampled and then completed using a low-rank matrix recovery method, where $\mb X_{\rm{L}}\in\mathbb{C}^{N_r\times d_1}$ and $\mb X_{\rm{R}}\in\mathbb{C}^{N_t\times d_2}$ are feature matrices. Clearly, when $d_1=N_r, d_2=N_t$, $\mb H$ can then be recovered as 
\[  {\mb H}  = {(\mb X^H_{\rm{L}}})^{-1}  {\mb C}(\mb X_{\rm{R}})^{-1}\] 
when $\mb C$ is known. Obtaining the entries of $\mb C$ is equivalent to using the columns of $\mb X_{\rm{L}}$ and $\mb X_{\rm{R}}$ to sample $\mb H$, i.e., 
\begin{equation}
\label{IMC_sample}
[\mb C]_{i,j}=\mb X^H_{\rm{L}}(:,i) \mb H\mb X_{\rm{R}}(:,j).
\end{equation}
Therefore, when the feature matrices $\mb X_{\rm{L}}$ and $\mb X_{\rm{R}}$ are known, the sampling process can be achieved by setting the precoder $\mb f_m$  and combiners  $\mb W_{m,s}$ of (\ref{y_ts}) as columns of $\mb X_{\rm{R}}$ and $\mb X_{\rm{L}}$, respectively. As such, the numbers of antennas activated simultaneously are given by the numbers of nonzero elements in the columns of $\mb X_{\rm{L}}$ and $\mb X_{\rm{R}}$. Note that the IMC formulation here reduces to the MC approach when $\mb X_{\rm{L}}= \mb I_{N_r}, \mb X_{\rm{R}}=\mb I_{N_t}$. In the following, we focus on the choice of $\mb X_{\rm{L}}$ and $\mb X_{\rm{R}}$.

Let $\mb H=\mb U\mb S\mb V^H$ be the thin SVD of $\mb H$ with rank $r_{\rm{ch}}$, 
and let $\mb x_{{\rm{L}}_i}$ $(\mb x_{{\rm{R}}_i})$ be the $i$-th column of $\mb X_{\rm{L}}$ $(\mb X_{\rm{R}})$. In order to successfully recover $\mb C$ and $\mb H$, the feature matrices $\mb X_{\rm{L}}$ and $\mb X_{\rm{R}}$ have to satisfy the following two key properties \cite{RobustIMC}. 
\begin{enumerate}
\item \emph{Incoherent w.r.t $\mb H$}: The feature matrices $\mb X_{\rm{L}}$ and $\mb X_{\rm{R}}$ are incoherent with respect to $\mb H$, i.e.,
\begin{align}
&\max_{i}\|\mb U^H\mb x_{{\rm{L}}_{i}}\|_2\leq \sqrt{\frac{\mu_0 r_{\rm{ch}} }{N_r}},\\
&\max_{j}\|\mb V^H\mb x_{{\rm{R}}_{i}}\|_2\leq \sqrt{\frac{\mu_0r_{\rm{ch}}}{N_t}},\\
&\max_{i,j}\|\mb x^H_{{\rm{L}}_i}\mb U\mb V^H\mb x_{{\rm{R}}_i}\|_2\leq \sqrt{\frac{\mu_0r_{\rm{ch}}}{N_rN_t}}
\end{align}
\item \emph{Self-incoherent}: The feature matrices $\mb X_{\rm{L}}$ and $\mb X_{\rm{R}}$ are both $\mu_1$-incoherent, i.e.,
\begin{equation}
\label{self}
\max_{i}\|\mb x_{{\rm{L}}_{i}}\|_2\leq \sqrt{\frac{\mu_1d_1}{N_r}},\quad \max_{j}\|\mb x_{{\rm{R}}_{i}}\|_2\leq \sqrt{\frac{\mu_1d_2}{N_t}}
\end{equation}
\end{enumerate} 
The above properties imply that matrix $\mb C$ should not be too spiky so that it is possible to be recovered from a subset of entries \cite{matrix completion with noise}.  
Moreover, if $\mb X_{\rm{L}}$ and $\mb X_{\rm{R}}$ have orthonormal columns, i.e., $\mb X^H_{\rm{L}}\mb X_{\rm{L}}=\mb I_{N_r}$ and $\mb X^H_{\rm{R}}\mb X_{\rm{R}}=\mb I_{N_t}$, the condition number of $\mb C$ and that of $\mb H$ are equal. 
This is useful because if the condition numbers differ, a practical matrix completion algorithm may produce an estimate of $\mb C$ with a different rank. This can in turn yield over- or underestimation the rank of $\mb H$. However, not all the orthonormal matrices are suitable for $\mb X_{\rm{L}}$ and $\mb X_{\rm{R}}$. 
For example, 
consider an extreme case where the AoAs/AoDs 
coincide with the normalized spatial frequencies and $\mb X_{\rm{L}}$  and $\mb X_{\rm{R}}$ are unitary DFT matrices. Then 
it can be verified that the transformed matrix 
$\mb C$ becomes a diagonal matrix, which is sparse and very spiky and can hardly be recovered unless all of its entries are observed \cite{matrix completion with noise}. 

In light of the above discussion, we choose $\mb X_{\rm{L}}$ and $\mb X_{\rm{R}}$ as follows:
\begin{itemize}
\item Obtaining two matrices $\mb A\in\mathbb{C}^{N_r\times N_r}$ and $\mb B\in\mathbb{C}^{N_t\times N_t}$ whose elements are generated randomly on a unit circle. 
\item Calculate the SVD of $\mb A$ and $\mb B$ as $\mb A=\mb U_{\rm{A}}\mb S_{\rm{A}}\mb V^H_{\rm{A}}$ and $\mb B=\mb U_{\rm{B}}\mb S_{\rm{B}}\mb V^H_{\rm{B}}$.
\item Set $\mb X_{\rm{L}}=\mb U_{\rm{A}}$ and $\mb X_{\rm{R}}=\mb U_{\rm{B}}$.
\end{itemize}
With $\mb X_{\rm{L}}$ and $\mb X_{\rm{R}}$ given, noisy observations of a subset of the entries of $\mb C$ are obtained by choosing $\mb f_m$ and $\mb W_{m,s}$ of (\ref{y_ts}) as the corresponding column(s) of $\mb X_{\rm{R}}$ and $\mb X_{\rm{L}}$, respectively. For example, in order to observe $[\mb C]_{1,1}$ and $[\mb C]_{2,1}$ at the $s$-th step of the $m$-th training stage, 
we can set $\mb W_{m,s}=\mb X_{\rm{L}}(:,1:2)$ and $\mb f_{m}=\mb X_{\rm{R}}(:,1)$ and obtain  
\begin{equation}
\label{WNM}
\begin{bmatrix}
[\widetilde{\mb C}]_{1,1}\\
[\widetilde{\mb C}]_{2,1}
\end{bmatrix}=\mb W^H_{m,s}\mb H\mb f_{m} s_{m,s} +\mb W^H_{m,s}\mb n_{m,s},
\end{equation}
where $\mb n_{m,s}$ denotes the observation noise. 
Note that (\ref{WNM}) is actually the same as (\ref{y_ts}). The corresponding $\rm PNR$ can be defined in the same way as (\ref{PNR}).  
We choose the sampling domain $\Omega$ the same as in Section \ref{training}, which takes $N/N_t$ distinct noisy samples from the $N_r$ entries of each column of $\mb C$. 
Note that $\mb f_m$ and $\mb W_{m,s}$ in (\ref{WNM}) are no longer made of only $1$'s and $0$'s, and thus the design discussed in Section \ref{training} is not suitable here. We adopt the PE-AltMin algorithm in \cite{AltMin} to solve (\ref{design task F}) and (\ref{design task W}) for realizing $\mb f_m$ and $\mb W_{m,s}$ using the hybrid transceivers.  

Similar to (\ref{Y}), after $MS$ training steps, we obtain the received samples in $\mb Y_{\rm{C}}$ and then construct a matrix $\widetilde{\mb C}\in\mathbb{C}^{N_r\times N_t}$ as
\begin{equation}
\label{C_tilde}
[\widetilde{\mb C}]_{l,k}=
\begin{cases}
[\mb Y_{\rm{C}}]_{i_{l,k},j_{l,k}}, &(l,k)\in\Omega,\\
& i_{l,k}=1,\ldots,\frac{N}{M}, j_{l,k}=1,\ldots,M \\
0, & \text{otherwise,}
\end{cases}
\end{equation}
Then matrix $\mb C$ can be estimated by solving the low-rank matrix recovery problem 

\begin{equation} 
\label{nyMCs}
\min_{\widehat{\mb C}} {\rm rank}(\widehat{\mb C}), \quad \quad \mathrm{s.t.} \quad \|P_{\Omega}( \widehat{\mb C})-P_{\Omega}(\widetilde{\mb C})\|^2_F\leq \delta^2_c,
\end{equation}
where $\delta^2_c$ is set according to the noise variance. 
Our proposed GCG-Alt estimator in Algorithm 1 can be directly applied to solve (\ref{nyMCs}) and has the same computational complexity as analyzed in Table 1. After obtaining $\widehat{\mb C}$, we can produce the estimate of the original channel matrix as \[ \widehat{\mb H}={(\mb X^H_{\rm{L}}})^{-1} \widehat{\mb C} (\mb X_{\rm{R}})^{-1}.\] This IMC formulation is still immune to the phase/gain errors as no knowledge of the array response is needed.

Note that \cite{MC_CS} also adopts the formulation of $\mb C=\mb X^H_{\rm{L}}\mb H\mb X_{\rm{R}}$ with the entries of $\mb X_{\rm{L}}$ and $\mb X_{\rm{R}}$ randomly generated from a unit circle, but it does not require the columns of $\mb X_{\rm{L}}$ and $\mb X_{\rm{R}}$  to be orthonormal. Therefore, the condition number of $\mb C$ may differ from that of $\mb H$ and the recovery accuracy may be affected. In addition, \cite{MC_CS} chooses $d_1< N_r$ and $d_2<N_t$ so that the dimension of $\mb C$ is smaller than $\mb H$, yielding lower computational complexity for the MC algorithms. However, after obtaining $\widehat{\mb C}$, \cite{MC_CS} needs to solve a CS problem, which requires the knowledge of the array response, to recover $\mb H$ from $\widehat{\mb C}$. As analyzed in Section I, if the array response is not accurately known, the performance of the CS solvers can degrade.

\section{Numerical Results}
We now evaluate the performance and computational complexity of our proposed design for fully connected hybrid transceivers with the ULA and USPA. 

\subsection{The ULA System}

We assume a carrier frequency of $f_c=28$ GHz. The number of clusters $K\sim \max({\rm Poisson}(1.8),1)$, and the cluster powers are generated following \cite[Table I]{measurements}. The number of rays in each cluster $L\sim\mathcal{U}[1,20]$. The horizontal AoDs 
\[ \phi^t_{kl} \sim \mathcal{U}(\phi^t_k-\upsilon^t_h/2,\phi^t_k+\upsilon^t_h/2),\]
where the center angles $\phi^t_k$ are distributed uniformly from $[0,2\pi]$ and separated by at least one angular spread $\upsilon^t_{h}=10.2^{\degree}$. 
 Similarly, the horizontal AoAs 
\[ 
\phi^r_{kl}\sim\mathcal{U}(\phi^r_k-\upsilon^r_h/2,\phi^r_{k}+\upsilon^r_h/2)
\]
with $\upsilon^r_h=15.5^{\degree}$. 
The noise is assumed to be additive white Gaussian noise (AWGN) with variance $\sigma^2$. 
The ULA at the BS has $N_t=128$ antennas and $K_t=16$ RF chains. 
The ULA at the MS has $N_t=32$ antennas and $K_r=4$ RF chains.
The RF beamformers employ $6$-bit phase shifters. Denote by $\varkappa^t$ and $\varkappa^r$ the phase error levels for the ULAs at the BS and MS, respectively. 
The phase errors of ULAs at the BS and MS are distributed respectively as   
\[ 
    \kappa^t_i\sim \mathcal{U}(-\varkappa^t, \varkappa^t), \quad \rm{and} \quad   
   \kappa^r_i\sim\mathcal{U}(-\varkappa^r, \varkappa^r).
\]
The gains of the antennas assumed to be
\[ 
\rho^t_i\sim \mathcal{U}(1-\varrho^t, 1+\varrho^t), \quad {\rm and} \quad  
\rho^r_i\sim\mathcal{U}(1-\varrho^r, 1+\varrho^r),\] respectively, for the BS and MS, where $\varrho^t$ and $\varrho^r$ are the unequal gain levels for the ULAs at the BS and the MS, respectively.  
When the arrays of the hybrid transceiver are perfectly calibrated, $\varkappa^t=\varkappa^r=0$ and $\varrho^t=\varrho^r=0$.

\begin{figure*}
\centering
\minipage[t]{0.49\columnwidth}
\label{Steps_compare}
\includegraphics[width=\columnwidth]{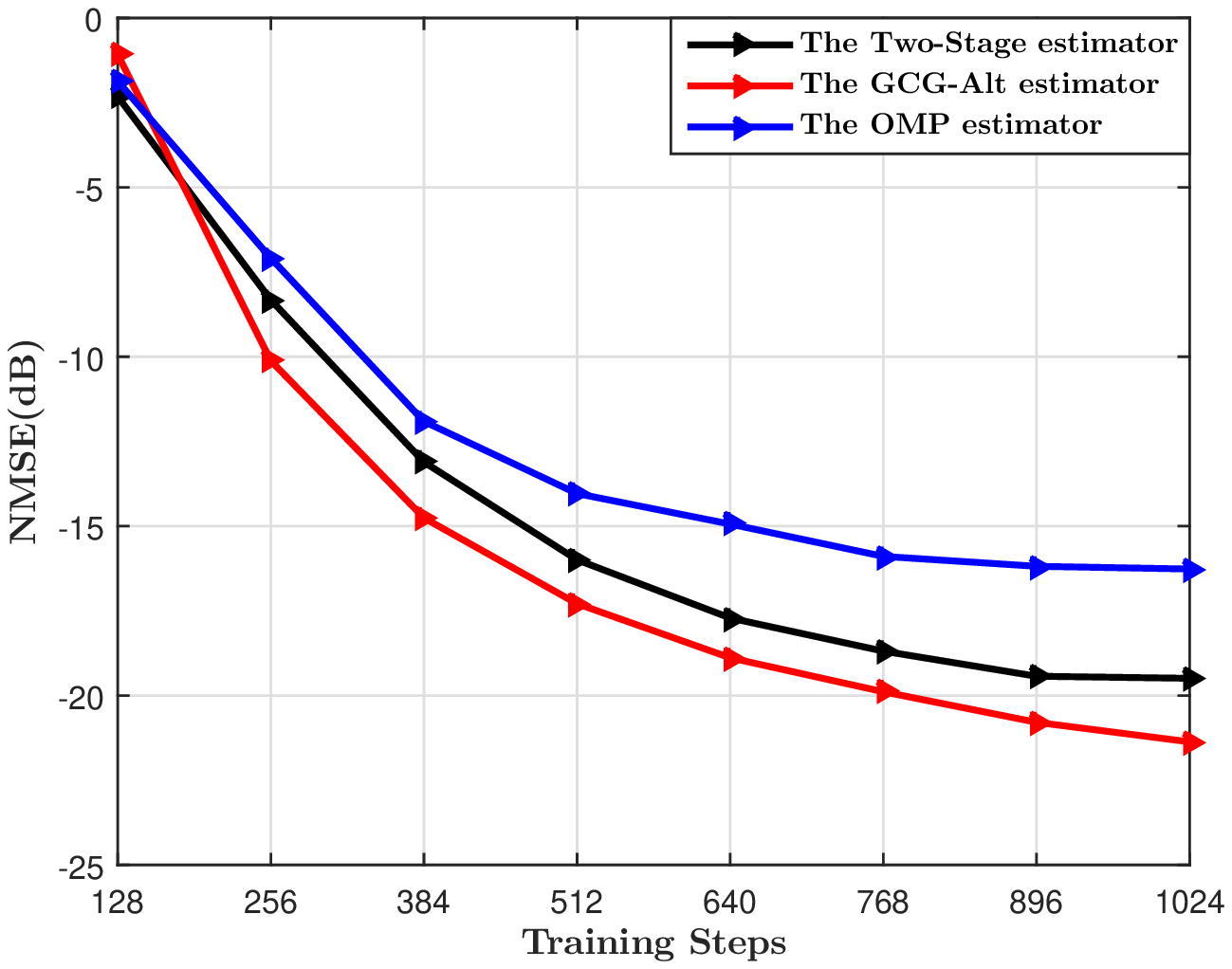}
\caption{NMSE of the channel estimation in the ULA system with $N_t=128, N_r=32, K_t=16, K_r=4$, different training steps, ${\rm{PNR}}=20$ dB, and perfectly calibrated arrays, i.e.,  $\varkappa^t=\varkappa^r=0, \varrho^t=\varrho^r=0$.}
\endminipage
\hfill
\minipage[t]{0.49\columnwidth}
\label{PNRs_compare}
\includegraphics[width=\columnwidth]{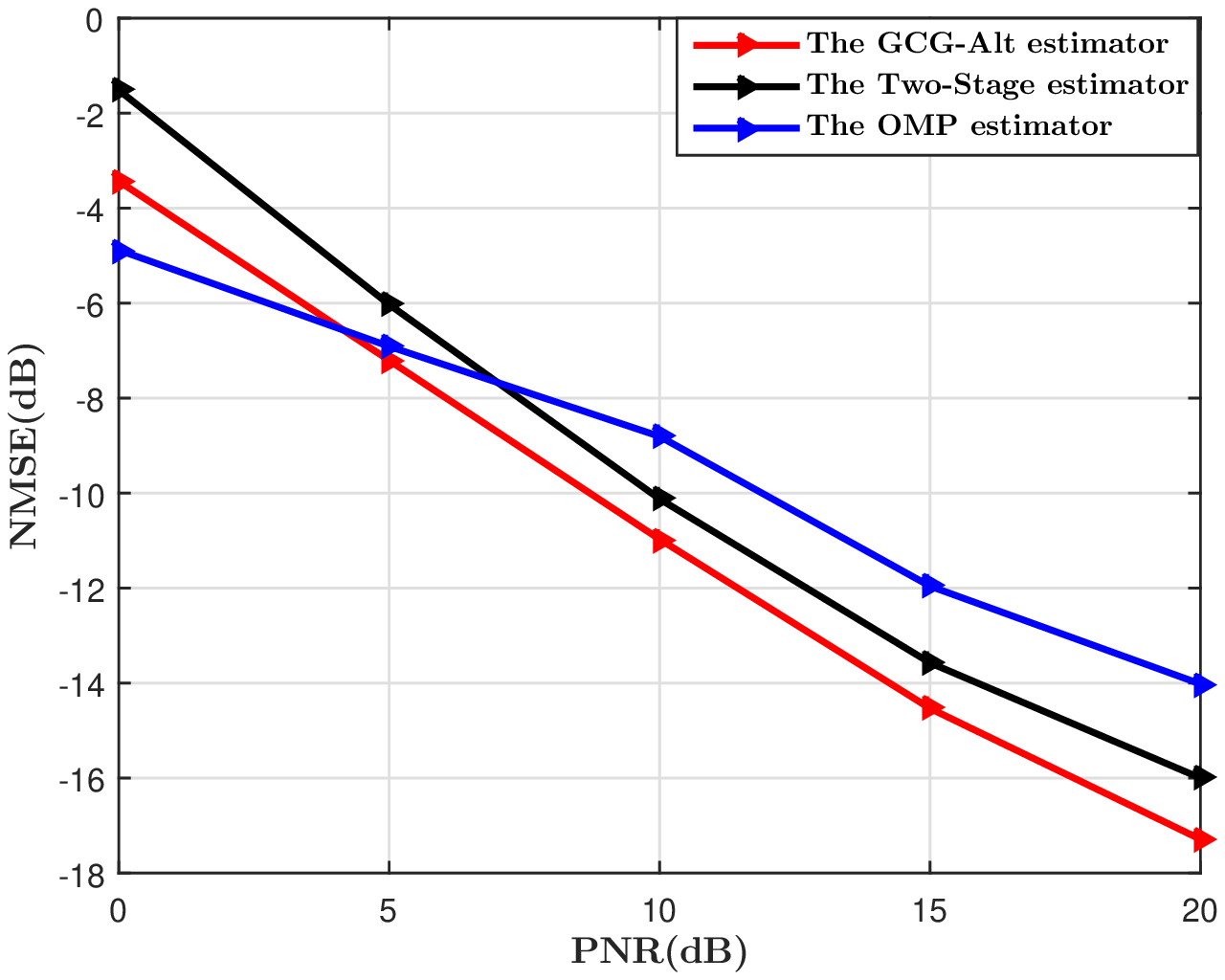}
\caption{NMSE of the channel estimation in the ULA system with $N_t=128, N_r=32, K_t=16, K_r=4$, $512$ training steps, different $\rm PNR$s and perfectly calibrated arrays, i.e.,  $\varkappa^t=\varkappa^r=0, \varrho^t=\varrho^r=0$.}
\endminipage
\end{figure*}

\begin{figure*}
\centering
\minipage[t]{0.49\columnwidth}
\label{Phaseerror_compare}
\includegraphics[width=1\columnwidth]{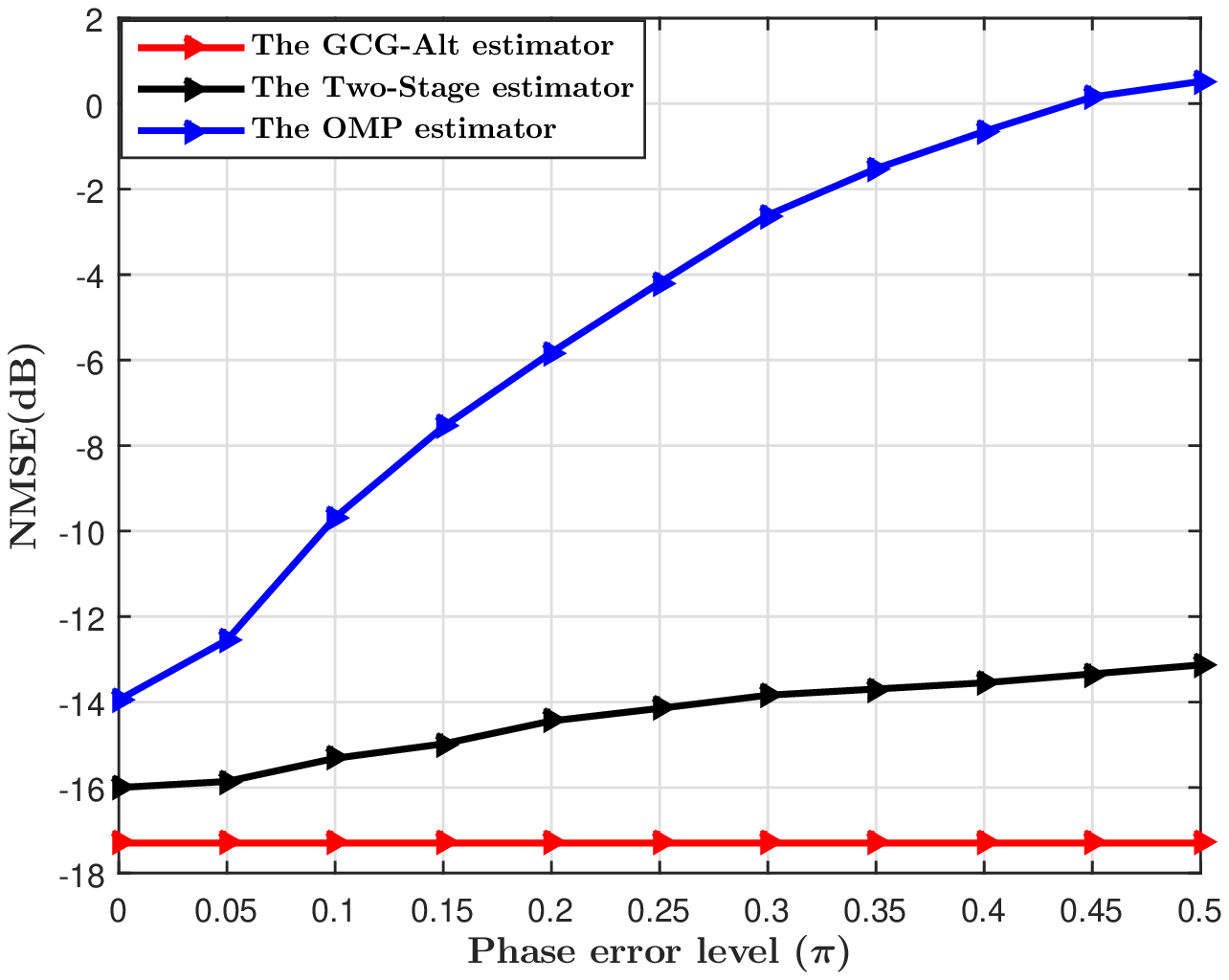}
\caption{NMSE of the channel estimation in the ULA system with $N_t=128, N_r=32, K_t=16, K_r=4$, $MS=512$ training steps, different phase error levels, ${\rm{PNR}}=20$ dB and $\varrho^t=\varrho^r=0$. The BS and MS phase error levels are assumed the same, i.e., $\varkappa^t=\varkappa^r$.}
\endminipage
\hfill
\minipage[t]{0.49\columnwidth}
\label{Gainerror_compare}
\includegraphics[width=1\columnwidth]{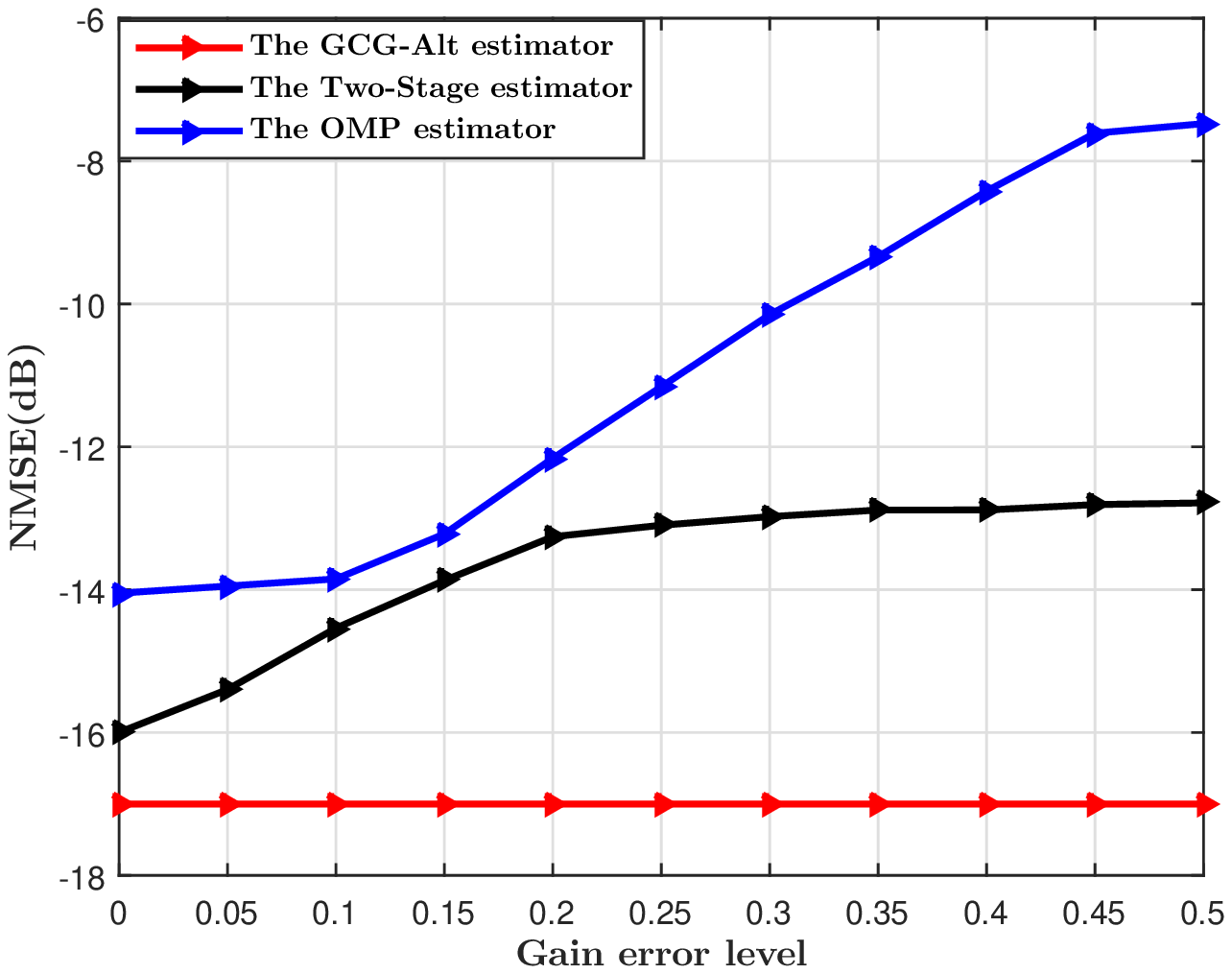}
\caption{NMSE of the channel estimation in the ULA system with $N_t=128, N_r=32, K_t=16, K_r=4$, $MS=512$ training steps, different gain error levels, ${\rm{PNR}}=20$ dB and $\varkappa^t=\varkappa^r=0$. The BS and MS gain error levels are assumed the same, i.e., $\varrho^t=\varrho^r$.}
\endminipage
\end{figure*}

 In this paper, we use the OMP estimator with the training beams optimized in \cite{channel estimation via OMP} to show the performance of the CS-based estimator. We choose the redundant dictionary with $G_t=2N_t=256$ and $G_{r}=2N_r=64$ for the OMP estimator. 
A stopping threshold $\epsilon_{\rm{OMP}}=0.1\sigma^2$ is set for the OMP estimator  in \cite{channel estimation via OMP}. 
Our observations show that for the present application, the stopping threshold is sensitive to the noise variance. For ${\rm{PNR}}<10$ dB, $\epsilon_{\rm{OMP}}=0.1\sigma^2$ leads to underestimation of the available paths, while for ${\rm{ PNR}}>10$ dB, $\epsilon_{\rm{OMP}}=0.1\sigma^2$ leads to overestimation and the OMP estimator takes too long to stop. 
 In order to show the potential of the OMP estimator, we set the optimized stopping threshold $\epsilon_{\rm{OMP}}= 0.025\sigma^2,  0.05\sigma^2,  0.1\sigma^2, 0.2\sigma^2, 0.4\sigma^2$ for ${\rm PNR}=0, 5, 10, 15, 20$ dB, respectively. Similar settings for the stopping threshold can be found in \cite{backhaul and access}. 
For our proposed GCG-Alt estimator, we set $\mu=\sigma^2, \epsilon=0.01, \epsilon_a=0.1$. We use our proposed training process and set $N_{m,s}=K_r-1=3$. Therefore, at each training step, the MS obtains $N_{m,s}=3$ samples. 
We also compare our proposed GCG-Alt estimator with the very recently proposed Two-Stage estimator in \cite{MC_CS}, which solves a MC problem using the FPC algorithm \cite{FPC} at the first stage and then solves a CS problem using FISTA \cite{FISTA} at the second stage. We notice that our proposed training scheme in Section \ref{training} outperforms the random training scheme in \cite{MC_CS} for the Two-Stage estimator for the channels considered in our simulations. We thus adopt our proposed training scheme when testing the Two-Stage estimator with the two design matrices $\mb Z$ and $\mb F$ of \cite{MC_CS} set as $\mb I_{N_r}$ and $\mb I_{N_t}$, respectively. The numbers of the BS and MS grid points for FISTA are $G^{\prime}_t=N_t=128$ and $G^{\prime}_r=N_r=32$, respectively.

We first assume the arrays of the hybrid transceiver are perfectly calibrated, i.e., $\varkappa^t=\varkappa^r=0$ and $\varrho^t=\varrho^r=0$. We compare the three estimators' performances under different training steps. The average of the normalized mean square error 
\[ 
{\rm NMSE}= \frac{\|\widehat{\mb H}-\mb H\|^2_F}{ \|\mb H\|^2_F}
\]
 is used to evaluate their performances, where $\widehat{\mb H}$ denotes the estimate of the channel matrix. 
For the OMP estimator in \cite{channel estimation via OMP}, the BS sends out $M$  transmitting beams and the MS uses $SK_r$ receiving beams for each transmitting beam to obtain a total of $MSK_r$ measurements in $MS$ training steps. For the GCG-Alt and the Two-Stage estimators, $MS$ training steps yield $MSN_{m,s}$ measurements.  
We fix $M=N_t=128$ for the three estimators, and set $S=1$ to $8$ training steps for each stage, yielding $128$ to $1024$ training steps in total. 
We set ${\rm PNR}=20$ dB, which may be feasible for some scenarios such as the backhaul and access links in ultra-dense networks \cite{backhaul and access}.
From Fig. 2, when the number of training steps is small, i.e., the sampling density $p$ is low, the Two-Stage estimator outperforms the GCG-Alt estimator and the OMP estimator. As the number of training steps increases, the performance for all three estimators improves and the GCG-Alt estimator performs the best. 
Fig. 3 shows the channel estimation performance with $MS=512$ training steps, which corresponds to a sampling ratio of $p=0.5$ for the OMP and $p=0.375$ for the GCG-Alt estimator and the Two-Stage estimator. Different $\rm PNR$s are considered. 
The results suggest that the GCG-Alt estimator has better recovery performance  when ${\rm PNR}\geq 5$ dB. 

We also consider imperfectly calibrated BS and MS arrays.
 Fig. 4 and 5 compare the performance with different levels of phase and gain errors. It is seen that the performance of the GCG-Alt estimator remains stable while the performance of the OMP estimator and the Two-Stage estimator degrades as the phase or gain error level increases. The performance deterioration of the Two-Stage estimator comes from its second stage where a CS method requiring the knowledge of the array response is applied. Thus, when the phase or gain errors are present, channel estimators relying on the knowledge of the array response may suffer from performance degradations. 

\begin{figure*}[!t]
		 \centering{\subfloat[The distribution of $r_{\rm{sub}}$ ]{\includegraphics[width=0.33\columnwidth]{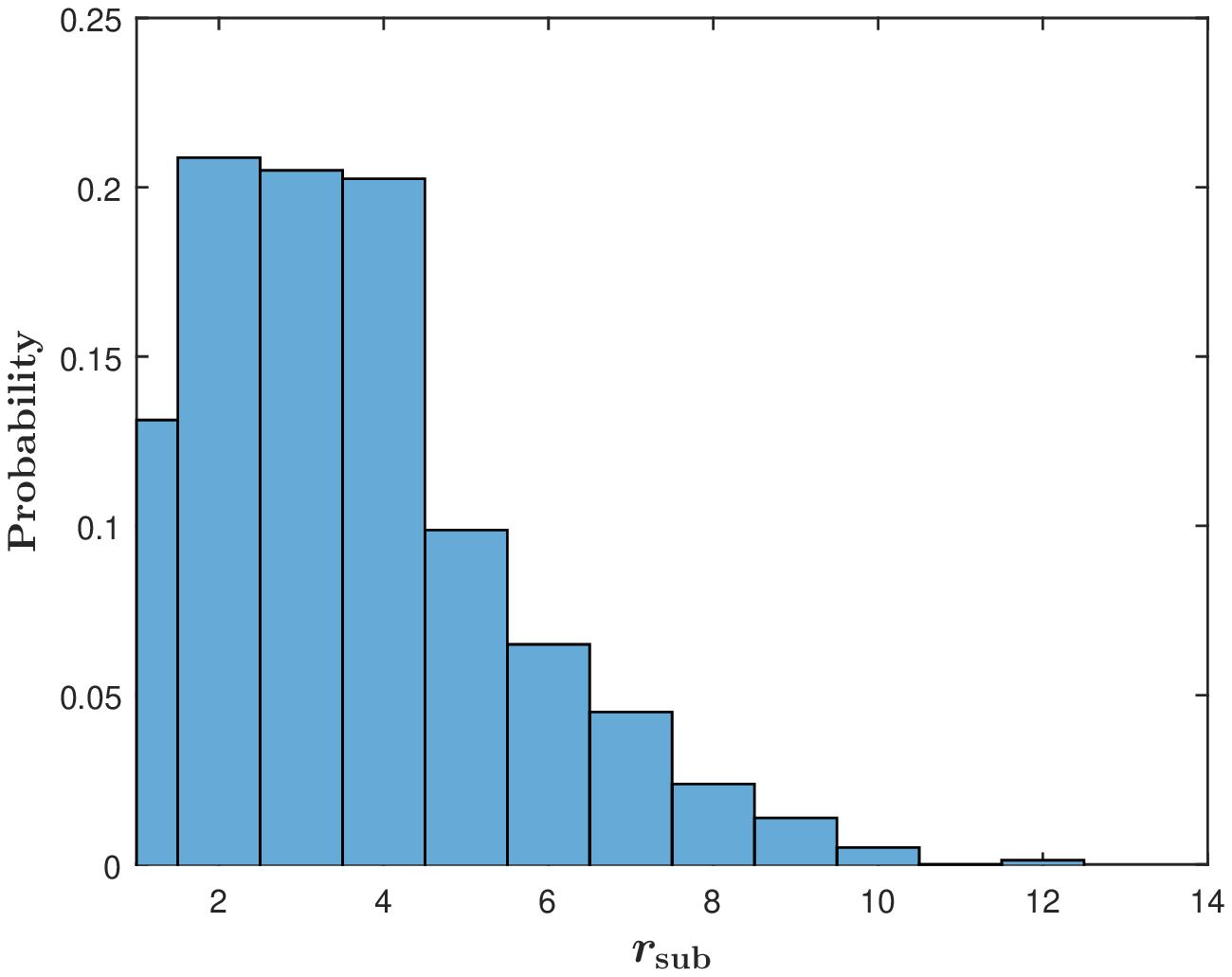}
		            \label{s}}	           
	            \subfloat[The distribution of $\widehat{r}_{\rm{GCG}}$.]{\includegraphics[width=0.33\columnwidth]{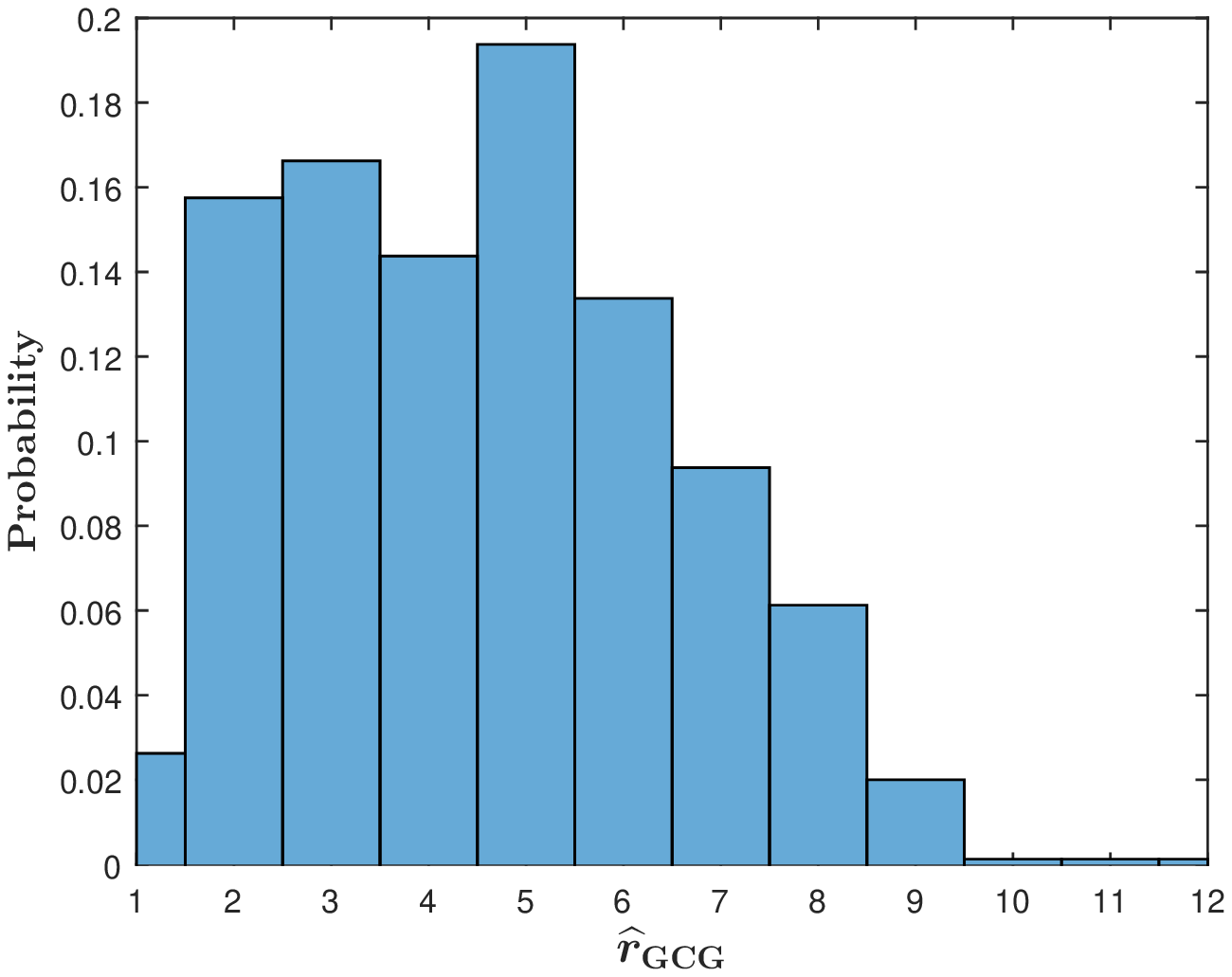}
	            \label{s}}
                      \subfloat[The distribution of $\widehat{r}_{\rm{OMP}}$.]{\includegraphics[width=0.33\columnwidth]{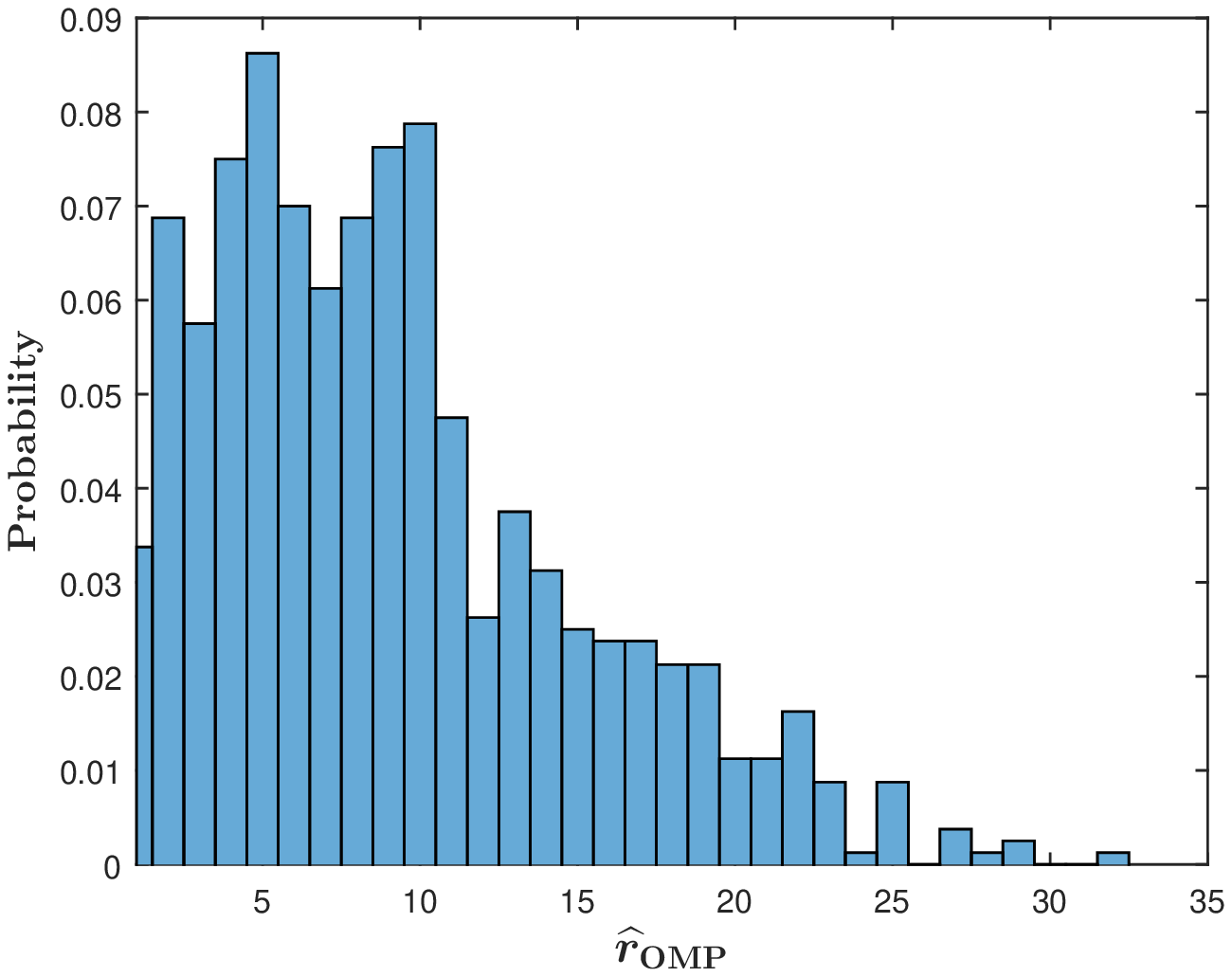}
	            \label{s}}}
	    \caption{Distributions of $r_{\rm{sub}}, \widehat{r}_{\rm{GCG}}$ and $\widehat{r}_{\rm{OMP}}$, with perfectly calibrated arrays ($\varkappa^t=\varkappa^r=0, \varrho^t=\varrho^r=0$), $MS=512$ training steps, and ${\rm  PNR}=20$ dB.}
	     \label{S}
\end{figure*}

\begin{figure*}
\centering
\minipage[t]{0.49\columnwidth}
\label{Iteration_change}
\includegraphics[width=\columnwidth]{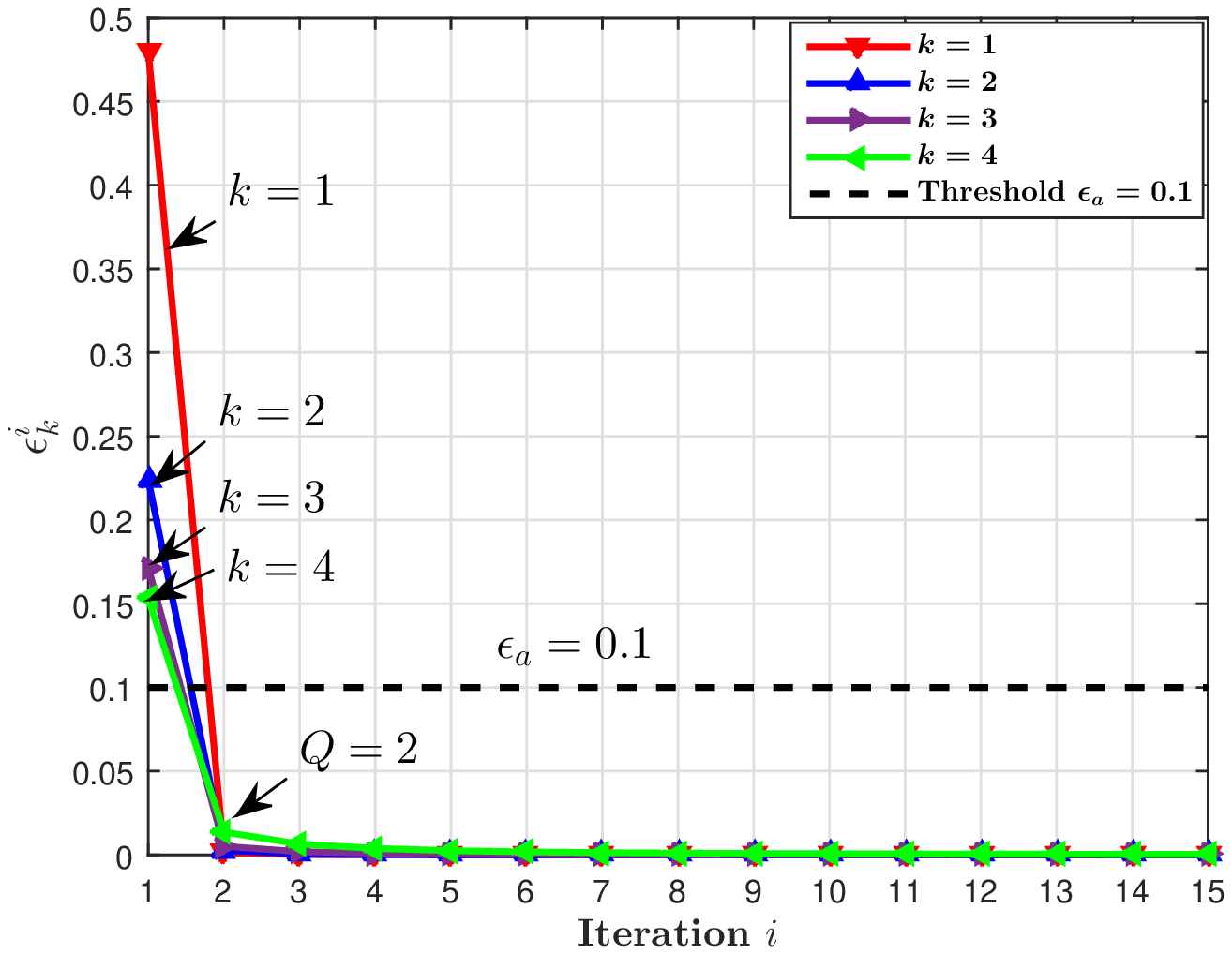}
\caption{Convergence rate of the AltMin algorithm with ${\rm{PNR}}=20$ dB, $MS=512$ training steps, and perfectly calibrated arrays, i.e.,  $\varkappa^t=\varkappa^r=0, \varrho^t=\varrho^r=0$.}
\endminipage
\hfill
\minipage[t]{0.49\columnwidth}
\label{Iteration_change}
\includegraphics[width=\columnwidth]{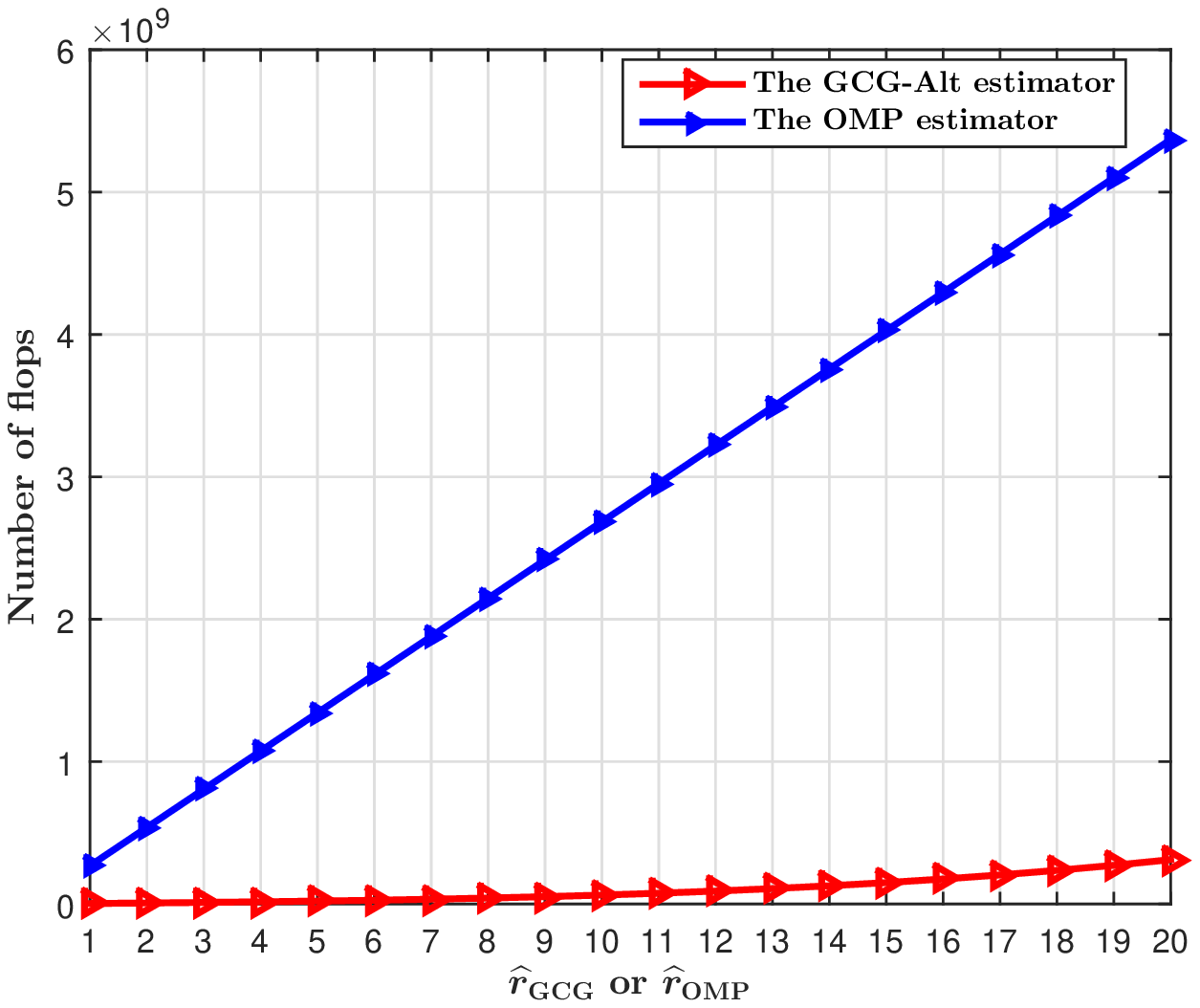}
\caption{Complexity comparison with different $\widehat{r}_{\rm{GCG}}$ (or $\widehat{r}_{\rm{OMP}}$), $N_t=128, N_r=32, Q=2$, $MS=512$ training steps. The parameters for the Randomized SVD method in the GCG algorithm are $q=2, g=3$, and the numbers of grid points of the redundant dictionary for the OMP estimator are $G_t=256 \text{ and }G_r=64$. }
\endminipage
\end{figure*}

We also examine the estimated rank of the channel using the OMP estimator and the GCG-Alt estimator.
We define $r_{\rm{sub}}$ as the rank of the reduced-rank approximation of the true channel that captures $95\%$ of the channel's energy and denote by $\widehat{r}_{\rm{GCG}}$ and $\widehat{r}_{\rm{OMP}}$ the ranks of the channel estimates produced by the GCG-Alt and OMP estimators,   respectively.
 The distribution of $r_{\rm{sub}},\widehat{r}_{\rm{GCG}}$ and $\widehat{r}_{\rm{OMP}}$ are illustrated in Fig. 6 for ${\rm PNR}=20$ dB. 
 From Fig. 6 (a), the probability of $r_{\rm{sub}}\leq 5$ is around $80\%$, and the probability of $r_{\rm{sub}}$ higher than $8$ is less than $5\%$. The distribution of $\widehat{r}_{\rm{GCG}}$ is similar to $r_{\rm{sub}}$. By contrast, the distribution of $\widehat{r}_{\rm{OMP}}$ has a longer tail, suggesting that the OMP estimator tends to overestimate the channel paths.

We next compare the computational complexity between the OMP estimator and the GCG-Alt estimator. The number of iterations of the GCG algorithm is equal to the estimated rank $\widehat{r}_{\rm{GCG}}$. The number of iterations of the AltMin algorithm $Q$ depends on the threshold $\epsilon_a$. 
Recall that at the $k$-th GCG iteration, the AltMin algorithm stops when $\epsilon^i_{k}\leq \epsilon_a$.
At ${\rm PNR}=20$ dB,  Fig. 7 illustrates an example showing how the value of $\epsilon^{i}_{k}$ changes over iterations for $k=1,2,3,4$. If $\epsilon_a=0.1$ is set, then the AltMin algorithm stops after $Q=2$ iterations. 
Based on the flop counts in Table 1, Fig. 8 shows the number of flops needed by the GCG-Alt estimator and the OMP estimator when we fix $Q=2$ and vary $\widehat{r}_{\rm{GCG}}$ or $\widehat{r}_{\rm{OMP}}$ from $1$ to $20$. 
Note that from Fig. 6, $\widehat{r}_{\rm{OMP}}$ tends to be larger than $\widehat{r}_{\rm{GCG}}$. Therefore, the computational complexity of the proposed GCG-Alt estimator is much lower than the OMP estimator. 

In order to investigate the influence of channel estimation on the achievable SE of the hybrid transceiver, we use the PE-AltMin hybrid precoder  proposed in \cite{AltMin}. The data transmission model \cite{AltMin} is
\begin{equation}
\label{SE_model}
\mb y=\mb D^H \mb Q^H \mb H \mb G \mb B \mb s+\mb D^H \mb Q^H\mb n,
\end{equation}
where $\mb D, \mb Q, \mb B,\mb G$ are the MS digital processor, MS RF beamformer, BS digital processor and BS RF beamformer, respectively, $\mb s\in\mathbb{C}^{N_s}$ is the symbol vector with $\mathbb{E}[\mb s\mb s^H]=\frac{1}{N_s}\mb I_{N_s}$, $N_s$ is the number of data streams, and $\mb n$ is the noise vector. 
The reason of using the PE-AltMin precoder is that it is immune to array-inherent impairments as it does not rely on the antenna array response, and has lower computational complexity compared to other hybrid precoders such as \cite{spatially precoding}. 
The signal-to-noise ratio $\rm{(SNR)}$ is defined as the ratio between the total transmitting signal power $|| \mb G \mb B \mb s||^2$ and the noise power.
We set $N_s=K_r=4$. The SE result for ${\rm{PNR}}=10$ dB with perfectly calibrated BS and MS arrays is shown in Fig. 9. All of the three estimators can obtain the CSI that leads to near-optimal SE for $\rm{SNR}\leq 0$ dB, but the CSI provided by the Two-Stage estimator and the OMP estimator incurs higher SE loss than that provided by the GCG-Alt estimator when $\rm{SNR}>0$ dB.
   
When the arrays are not perfectly calibrated, e.g., with the phase error levels $\varkappa^t=\varkappa^r=0.25\pi$ and gain error levels $\varrho^t=\varrho^r=0.2$, the SE evaluation result is demonstrated in Fig. 10. The GCG-Alt estimator still provides relatively more accurate CSI, 
leading to higher SE. Moreover, since the Two-Stage estimator is less sensitive to the phase/gain errors, its SE loss compared to the OMP estimator is lower.

\begin{figure}
\centering
\minipage[t]{0.49\columnwidth}
\label{Ideal_SEcompare}
\includegraphics[width=\columnwidth]{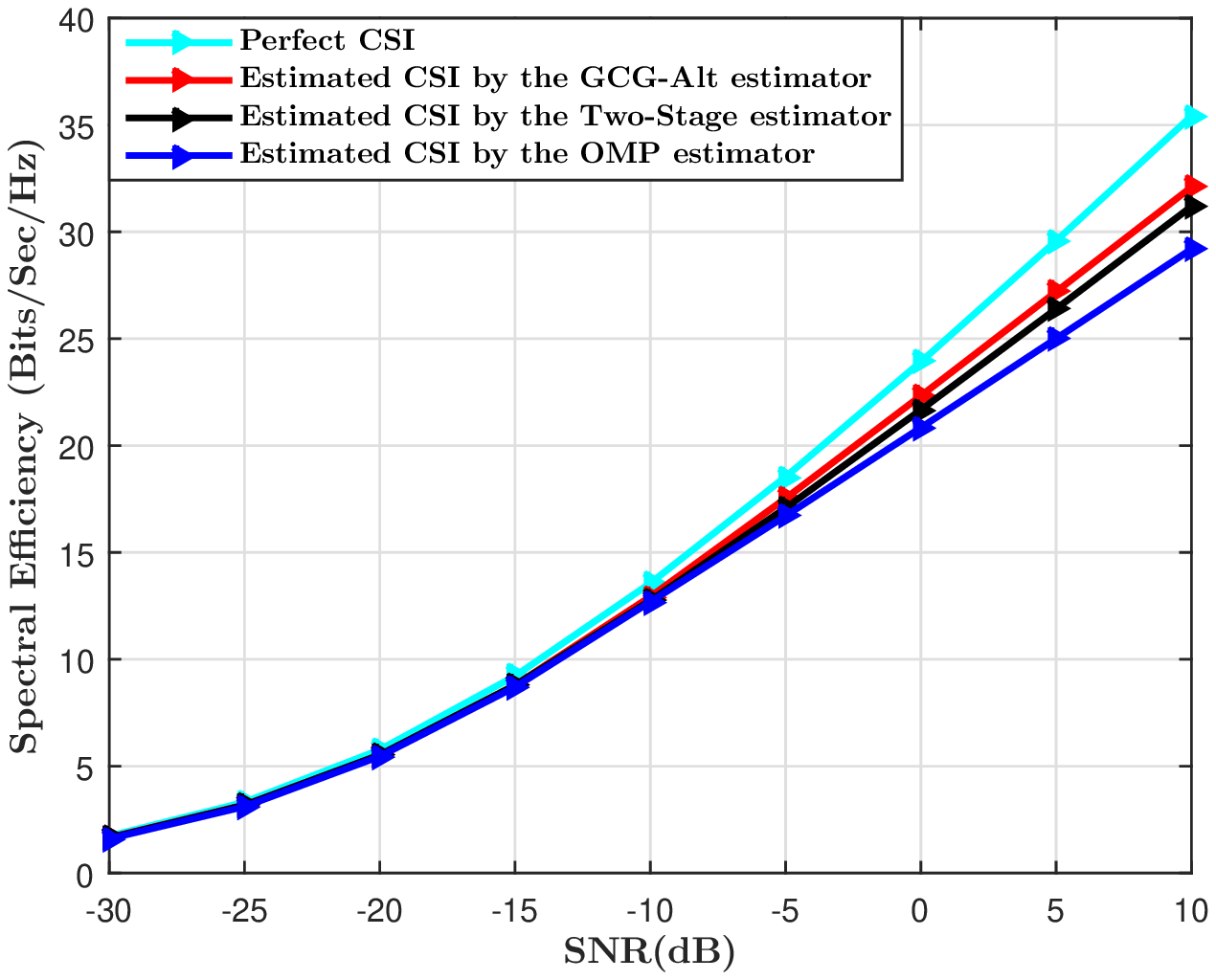}
\caption{Spectral efficiency achievable with different channel estimation schemes and the PE-AltMin precoder for the ULA system, $MS=512$ training steps, $N_s=4$, $\rm{PNR}=10$ dB, and perfectly calibrated arrays, i.e.,  $\varkappa^t=\varkappa^r=0, \varrho^t=\varrho^r=0$.}
\endminipage
\hfill
\minipage[t]{0.49\columnwidth}
\label{GainPhaseerror_compare}
\includegraphics[width=\columnwidth]{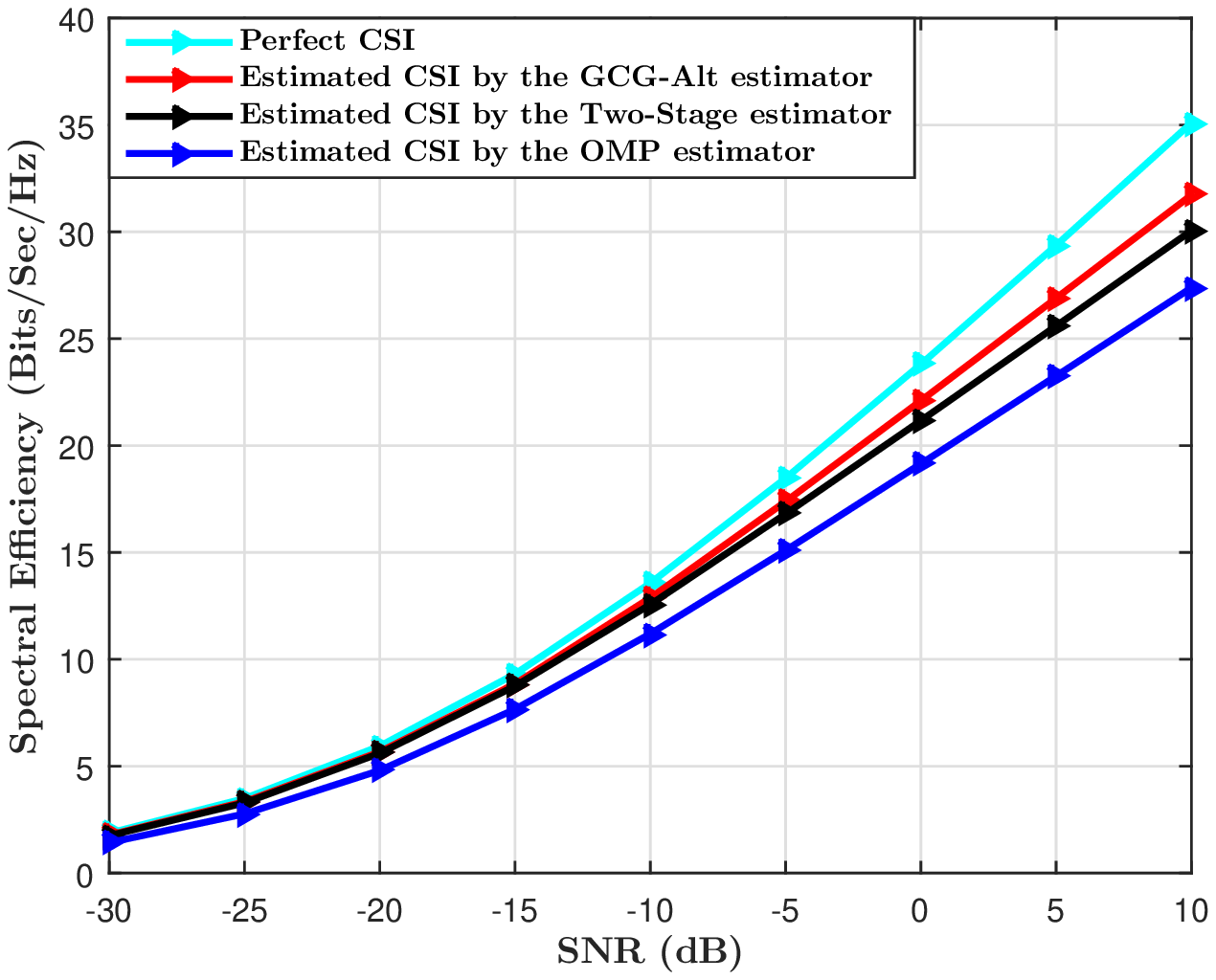}
\caption{Spectral efficiency achievable with different channel estimation schemes and the PE-AltMin precoder for the ULA system, $MS=512$ training steps, $N_s=4$, $\rm{PNR}=10$ dB, and imperfectly calibrated arrays with $\varkappa^t=\varkappa^r=0.25\pi, \varrho^t=\varrho^r=0.2$.}
\endminipage
\end{figure}


\begin{figure}
\centering
\minipage[t]{0.49\columnwidth}
\label{Ideal_SEcompare_USPA}
\includegraphics[width=\columnwidth]{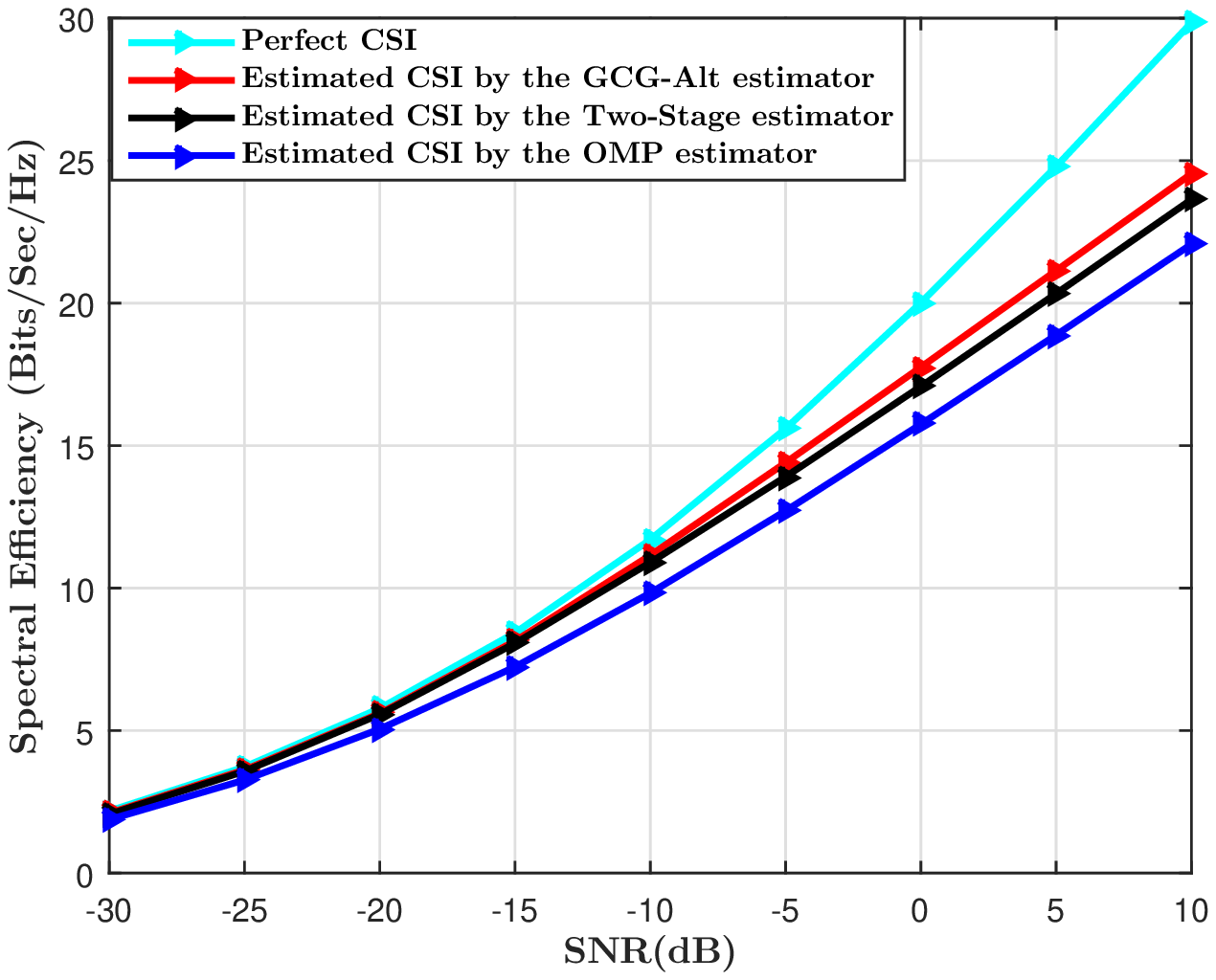}
\caption{Spectral efficiency achievable with different channel estimation schemes and the PE-AltMin precoder for the USPA system, $MS=576$ training steps, $N_s=4$, $\rm PNR =10$ dB, and perfectly calibrated arrays, i.e.,  $\varkappa^t=\varkappa^r=0, \varrho^t=\varrho^r=0$.}
\endminipage
\hfill
\minipage[t]{0.49\columnwidth}
\label{Iteration_change_USPA}
\includegraphics[width=\columnwidth]{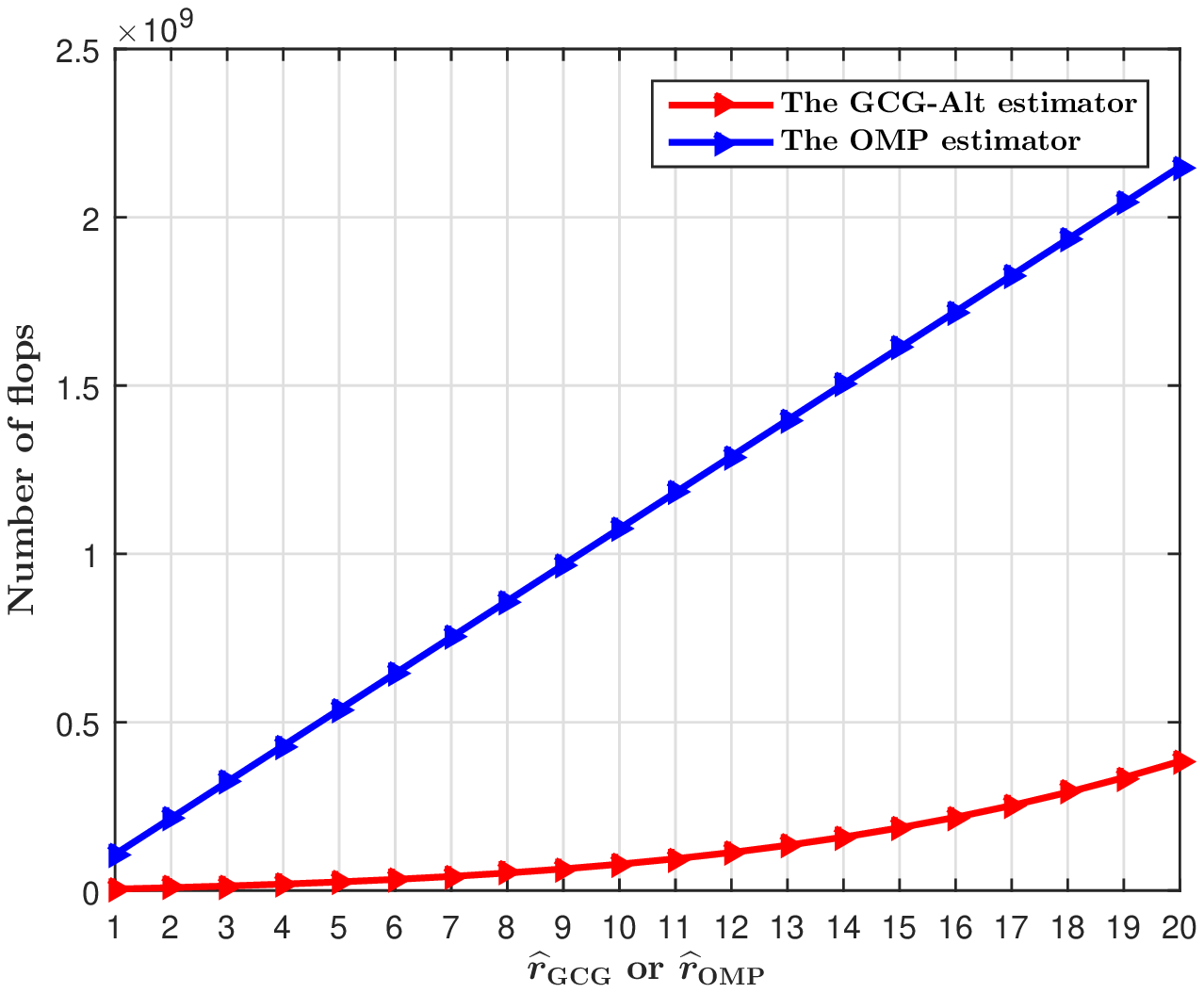}
\caption{Complexity comparison for the USPA system with different $\widehat{r}_{\rm{GCG}}$ (or $\widehat{r}_{\rm{OMP}}$), $N_t=144, N_r=36, Q=2$, $MS=512$ training steps. The parameters for the Randomized SVD method in the GCG algorithm are $q=2, g=3$, and the numbers of grid points of the unitary dictionary for the OMP estimator are $G_t=144 \text{ and } G_r=36$. }
\endminipage
\end{figure}


\begin{figure}
\minipage[t]{0.49\columnwidth}
\label{GainPhaseerror_compare_USPA}
\includegraphics[width=\columnwidth]{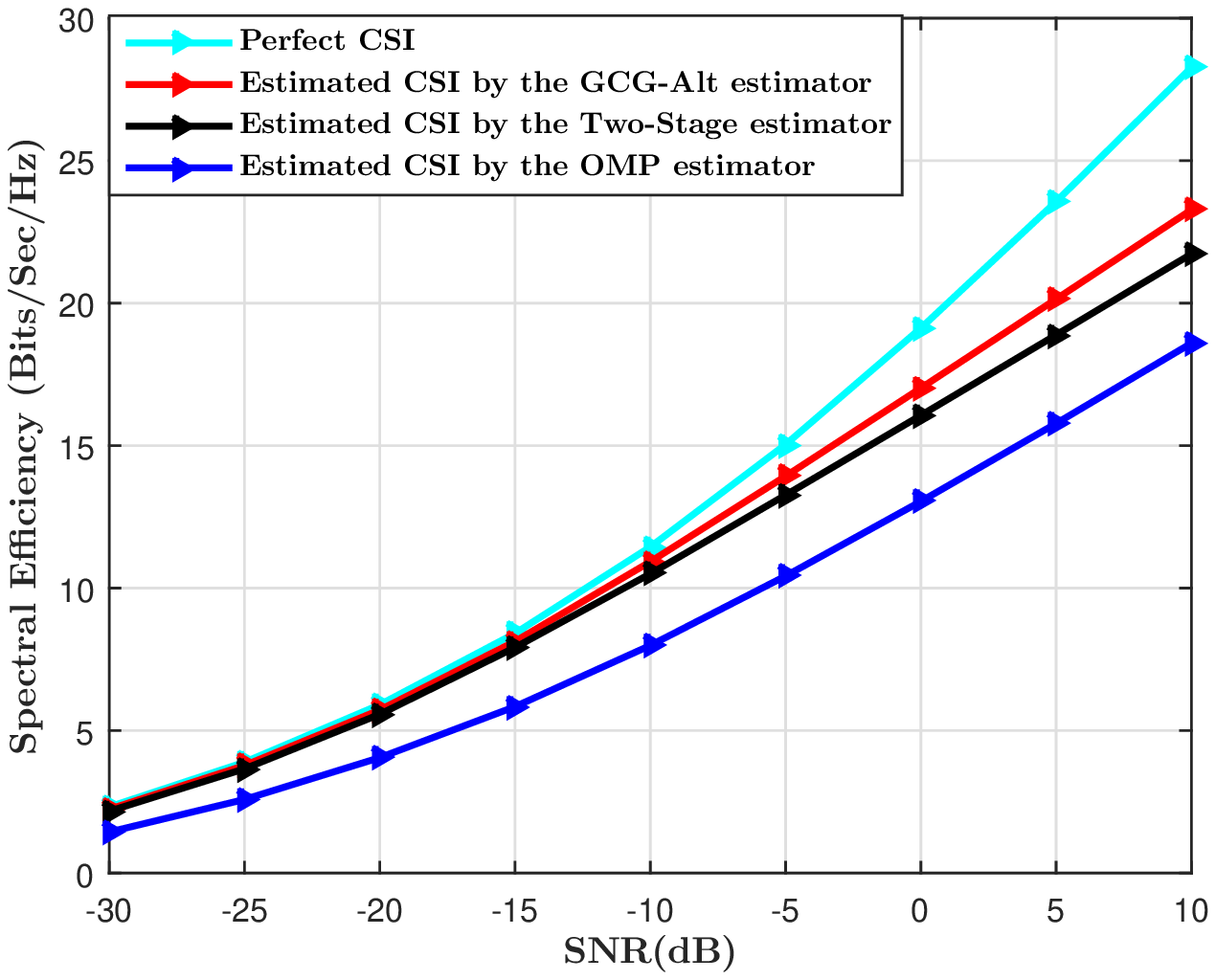}
\caption{Spectral efficiency achievable with different channel estimation schemes and the PE-AltMin precoder for the USPA system, $MS=576$ training steps, $N_s=4$, $\rm{PNR}=10$ dB, and imperfectly calibrated arrays with $\varkappa^t=\varkappa^r=0.25\pi, \varrho^t=\varrho^r=0.2$.}
\endminipage
\hfill
\minipage[t]{0.49\columnwidth}
\label{F14}
\includegraphics[width=\columnwidth]{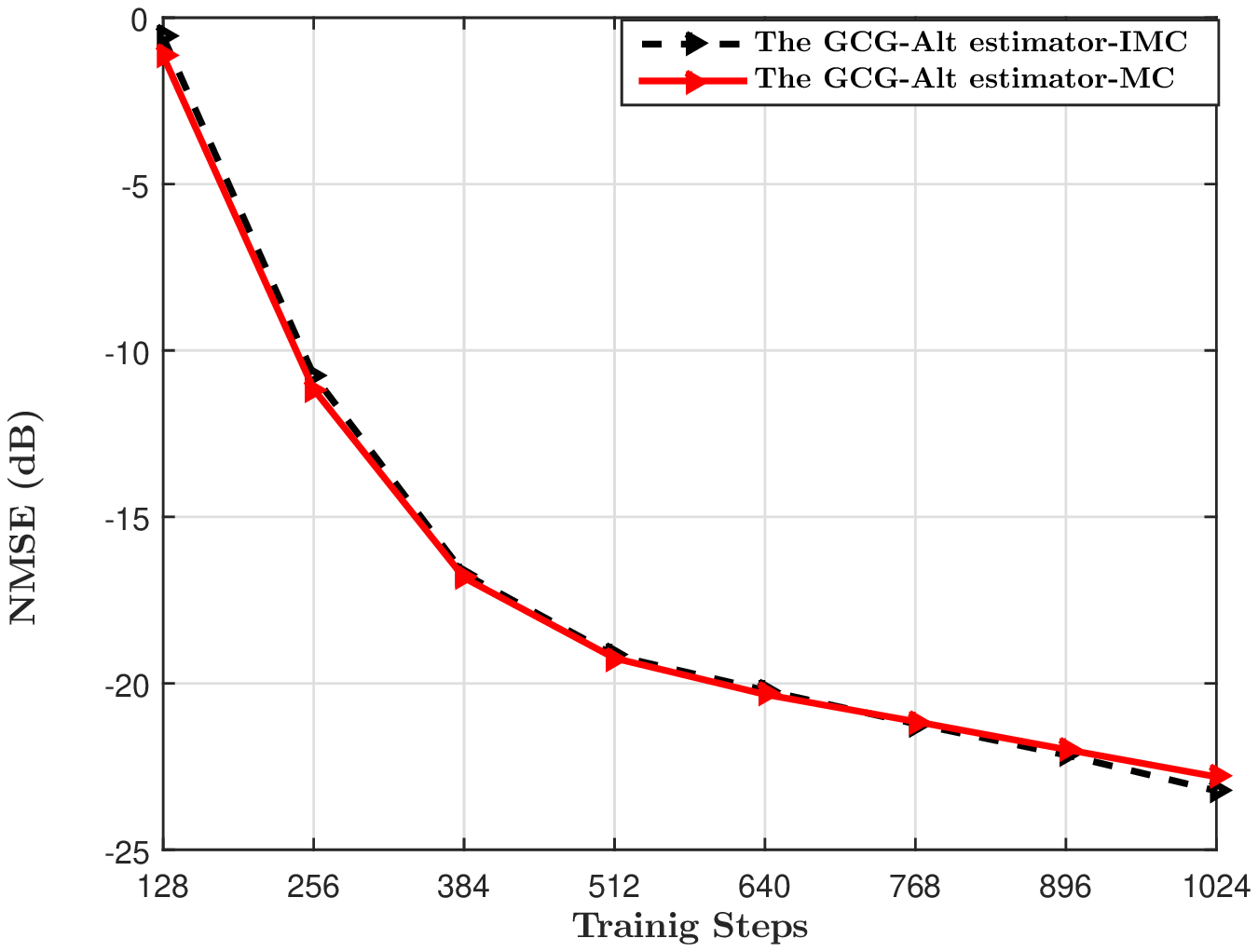}
\caption{NMSE of the channel estimation in the ULA system with $N_t=128, N_r=32, K_t=16, K_r=4$, different training steps, ${\rm{PNR}}=20$ dB, and perfectly calibrated arrays, i.e.,  $\varkappa^t=\varkappa^r=0, \varrho^t=\varrho^r=0$.}
\endminipage
\end{figure}


\subsection{ The USPA System} 

We next consider the system with USPA at the BS and MS. The parameters $f_c, K, L, \phi^{t}_{kl}, \phi^r_{kl}$ are assumed the same as in the ULA system. Based on the measurement results in \cite{measurements}, we assume the vertical AoD angular spread $\upsilon^t_v=0^{\degree}$ and the vertical AoA angular spread $\upsilon^r_v=6^{\degree}$. The vertical AoDs and AoAs are distributed as 
\[ 
     \theta^t_{kl}\sim\mathcal{U}(\theta^t_k-\upsilon^t_v/2,\theta^t_k+\upsilon^t_v/2), \quad 
 \theta^r_{kl}\sim\mathcal{U}(\theta^r_k-\upsilon^r_v/2,\theta^r_{k}+\upsilon^r_v/2) \]
 with the vertical center angles $\theta^t_{k}$ and $\theta^r_{k}$ being generated in the same manner as the horizontal center angles $\phi^t_{k}$ and $\phi^r_{k}$ in the ULA system. The USPA at the BS has $N_t=144$ antennas and $K_t=18$ RF chains. The USPA at the MS has $N_r=36$ antennas and $K_r=4$ RF chains. The phase error and gain error are the same as defined in the ULA system. 

In the USPA system, we use the unitary dictionary with $G_t=N_t \text{ and }G_r=N_r$ for the OMP estimator since the redundant dictionary takes too much storage space\footnote{For the USPA system with $N_t=12\times 12, N_r=6\times 6$, the redundant dictionary that doubles the grids along both axes ($y$ axis and $z$ axis) requires $G_t=576, G_r=144$. Therefore, the storage space needed by the redundant dictionary will be $5184\times 82944$.}.   
The parameters $\epsilon_{\rm{OMP}}, \epsilon_{a}, \epsilon$ and $\mu$ are the same as in the ULA system. The  number of training steps $MS=144\times 4=576$, leading to a sampling ratio of $p=0.5$ for the OMP and $0.375$ for the GCG-Alt estimator and the Two-Stage estimators.

We set the number of streams $N_s=4$ and $\rm{PNR}=10$ dB. 
The SE result with $\varrho^t=\varrho^r=0$ and $\varkappa^t=\varkappa^r=0$ shown in Fig. 11 suggests that using the CSI estimated by the OMP estimator has an obvious SE loss, which is caused by using the unitary dictionary that has lower resolution than the redundant dictionary.
The computational complexity comparison presented in Fig. 12 demonstrates that the proposed GCG-Alt estimator still has lower computational complexity than the OMP estimator with a unitary dictionary. 
The SE result with $\varkappa^t=\varkappa^r=0.25\pi$ and $\varrho^t=\varrho^r=0.2$ shown in Fig. 13 indicates that the GCG-Alt estimator still provides relatively more accurate CSI but the Two-Stage and OMP estimators suffer from array-inherent impairments and provide less accurate CSI, which is similar to the case of the ULA system.

\subsection{The IMC Formulation} 
In Section III-D, we have generalized the training scheme in Section \ref{training} and the channel is estimated using an IMC scheme. Assuming the same ULA system with perfectly calibrated arrays in Section IV-A, we compare the IMC scheme with the MC scheme introduced in Section \ref{training}.  The NMSE with different training steps is shown in Fig. 14. We can see that these two schemes have almost the same performance. The MC scheme in Section \ref{training} can be realized with very few bits phase shifters, e.g., $1$-bit phase shifters, yet the training scheme in Section III-D requires lower instantaneous power for the transmitter antennas. 

\section{Conclusions}

We have considered the impact of array-inherent impairments on the performance of the dictionary dependent  CS-based channel estimators for hybrid transceivers in mmWave communication systems. We show that array-inherent impairments can affect the array response, and thus degrades the performance of the CS-based estimators that utilize the array response to design dictionaries.  We propose an MC-based channel estimator that is independent of the array response to avoid the channel estimation error caused by imperfectly calibrating the antenna elements' phase centers and gains.
A training scheme and a  channel matrix recovery algorithm based on GCG and alternating minimization are designed. The numerical results show that our proposed MC-based channel estimator is robust against phase errors and gain errors of the antenna elements and has advantages over the CS-based estimators.


In the present work, narrowband systems are assumed. The proposed methods may be extended to wideband scenarios \cite{Wideband}, \cite{WidebandConf} in different manners. For example, they can be directly applied to the pilot subcarriers in an OFDM setting. They may also be combined with direction-finding methods such as the MUSIC \cite{MUSIC} to estimate the angles of the propagation paths. The property that different subcarriers may share the same AoAs/AoDs \cite{Wideband} may then be exploited to offer a good initial guess for the proposed GCG-Alt estimator or to reduce the solution space of CS-based estimators that aim to recover the paths information. In the case of uncalibrated arrays, direction finding methods that account for the unknown phase/gain errors, such as \cite{Spatial Signature} and \cite{Chi}, may be exploited to improve the robustness.  

\end{document}